\RequirePackage{ifpdf}
\documentclass[12pt,a4paper,hyper]{JHEP3}
\usepackage{epsfig}
\usepackage{amsmath,amsfonts,amssymb}
\usepackage{multirow}
\usepackage{slashed}
\usepackage{graphicx}
\usepackage{epsfig}
\usepackage{psfrag}
\usepackage{color}

\usepackage{ulem}
 \usepackage{slashed}

\def\={\; = \;}

\renewcommand{\Im}{\mbox{Im}}
\renewcommand{\Re}{\mbox{Re}}

\newcommand{\bem}{\begin{pmatrix}}
\newcommand{\eem}{\end{pmatrix}}

\def\h{\eta}

\def\be{\begin{equation}}
\def\ee{\end{equation}}
\def\ba{\begin{align}}
\def\ea{\end{align}}
\def\bse{\begin{subequations}}
\def\ese{\end{subequations}}
\def\1F1{{}_1\!F_1}
\def\2F0{{}_2\!F_0}

\def\h3{$\textrm{H}_3^+$}
\font\manual=manfnt
\def\dbend{\lower3.5pt\hbox{\manual\char127}}

\def\bar{\overline}

\def\CN{{\cal N}}

\def\rt2{\sqrt{2}}
\def\irt2{{1\over\sqrt{2}}}

\font\cmss=cmss10
\font\cmsss=cmss10 at 7pt

\def\IL{\relax{\rm I\kern-.18em L}}
\def\IH{\relax{\rm I\kern-.18em H}}
\def\rlx{\relax\leavevmode}
\def\ZZ{\rlx\leavevmode\ifmmode\mathchoice{\hbox{\cmss Z\kern-.4em Z}}
{\hbox{\cmss Z\kern-.4em Z}}{\lower.9pt\hbox{\cmsss Z\kern-.36em Z}}
{\lower1.2pt\hbox{\cmsss Z\kern-.36em Z}}\else{\cmss Z\kern-.4em
Z}\fi}
\newcommand{\beq}{\begin{equation}}
\newcommand{\eeq}{\end{equation}}
\newcommand{\bal}{\begin{equation}\begin{aligned}}
\newcommand{\eal}{\end{aligned} \end{equation}}
\newcommand{\bea}{\begin{eqnarray}}
\newcommand{\eea}{\end{eqnarray}}

\title{\center{Exact holography and black hole entropy \\ in $\mathcal{N}=8$ and $\mathcal{N}=4$ string theory }}

\preprint{}
\author{ Jo\~ao Gomes\\
\it Institute of Physics, University of Amsterdam, Science Park 904, 1098 XH Amsterdam\\
    Institute for Theoretical Physics, University of Utrecht, Leuvenlaan 3584 CE Utrecht\\

Email:
\email{J.M.VieiraGomes at uva.nl}
}

\abstract{We compute the exact entropy of one-eighth and one-quarter BPS black holes in $\mathcal{N}=8$ and $\mathcal{N}=4$ string theory respectively. This includes all the $\mathcal{N}=4$ CHL models in both $K3$ and $T^4$ compactifications. The main result is a measure for the finite dimensional integral that one obtains after localization of supergravity on $AdS_2\times S^2$. This measure is determined entirely by an anomaly in supersymmetric Chern-Simons theory on local $\text{AdS}_3$ and takes into account the contribution from all the supergravity multiplets. In Chern-Simons theory on compact manifolds, this is the anomaly that computes a certain one-loop dependence on the volume of the manifold. For one-eighth BPS black holes, our results are a first principles derivation of a measure proposed in arXiv:1111.1161, while in the case of one-quarter BPS black holes our result computes exactly all the perturbative or area corrections. Moreover, we argue that instantonic contributions can be incorporated and give evidence by computing the measure, which matches precisely the microscopics. Along with this, we find a unitary condition that truncates the answer to a finite sum of instantons in perfect agreement with a microscopic formula. Our results therefore solve a number of puzzles related to localization in supergravity and constitute a larger number of examples where holography can be shown to hold exactly.}


\begin{document}
\section{Introduction}

In recent years there has been remarkable progress in computing quantum corrections to the entropy of extremal black holes. By relating the exact entropy to the partition function of string fields in the near horizon geometry, the $\text{AdS}_2$ proposal of \cite{Sen:2008vm} has led to a novel insight into the holographic nature of these black holes. Recently, in the context of supersymmetric black holes, localization techniques in supergravity \cite{Dabholkar:2011ec,Dabholkar:2010uh,Dabholkar:2014ema} have opened the possibility of computing the $\text{AdS}_2$ partition function exactly for any value of the charges, thus constituting a remarkable step forward in our understanding of quantum corrections in gravity and more generally of finite $N$ corrections in holography.

In this work, we continue the study of the partition function $Z_{AdS_2}$ using localization techniques. Our goal is to  derive the exact measure for the finite dimensional integral that one obtains using localization \cite{Dabholkar:2010uh}- it has been a long-standing problem determining the exact contribution from all the supergravity multiplets. In particular we are interested in giving a fundamental principles derivation of the measure proposed in \cite{Dabholkar:2011ec} for the case of one-eighth BPS states in $\CN=8$ string theory and extend it also to one-quarter BPS states in $\CN=4$ string theory, that is, IIB on $K3\times T^2$ and CHL models including $T^4$ orbifolds. 
 
The reason to study  one-eighth and one-quarter BPS black holes is twofold. On one hand, from microscopics, the spectrum of BPS states, which is known exactly for a large class of charge configurations, includes rich information about non-perturbative physics. On the other hand, on the black hole side, we can use an {\it{equality between index and degeneracy }}\cite{Sen:2009vz,Dabholkar:2010rm} to extract the exact degeneracy $d(q)$ of the black hole that we can use to guide and test the bulk computations. 

From the microscopic study, we find that perturbative corrections to the area formula  are captured exactly by a modified Bessel function of the first kind\footnote{Here $I_{\nu}(z)$ is the modified Bessel function of the first kind which is defined as
\begin{equation}
I_{\nu}(z)=\frac{1}{2\pi i}\int_{\epsilon-i\infty}^{\epsilon+i\infty}\frac{dt}{t^{\nu+1}}\exp{\left[t+\frac{z^2}{4t}\right]},\;\epsilon>0
\end{equation}}, which in the case of one-eighth BPS states has the form \cite{Dabholkar:2011ec}
\begin{equation}
d(q)\simeq I_{7/2}(\pi\sqrt{Q^2P^2}),\qquad
\CN=8,\label{Bessel n=8}
\end{equation}
while for one-quarter BPS states it is 
\begin{equation}
d(q)\simeq \frac{P^2+4n_p}{\sqrt{P^2}}(P^2+8n_p)^{k+3/2}\;I_{k+3/2}\left(\pi \sqrt{Q^2(P^2+8n_p)}\right),\qquad \CN=4\label{Bessel n=4}
\end{equation}
with the constant $n_p=0,1$ for the CHL models on $T^4$ and $K3$ respectively and $k$ is a certain positive integer that depends on the CHL orbifold. Without loss of generality, we have set $Q.P=0$, where $Q^2,P^2$ and $Q.P$ are the T-duality invariants.
 The validity of the expressions (\ref{Bessel n=8}) and (\ref{Bessel n=4}) holds up to exponentially suppressed terms for sufficiently large charges. 

Formula (\ref{Bessel n=4}) was derived in \cite{Dabholkar:2010rm} for $k=10$, that is, for the case of one-quarter BPS states in IIB on $K3\times T^2$, which agrees with a Rademacher expansion  \cite{Dabholkar:2012nd}. In section \S\ref{micro N=4 sec} we present a novel way to compute the leading (\ref{Bessel n=4}) and subleading Bessel behaviour (\ref{intro inst bessels}) of the one-quarter BPS degeneracy for the CHL models on $T^4$ and $K3$. 

To next leading order, the $\CN=4$ formula (\ref{Bessel n=4}) is corrected by terms that are also of Bessel type. Developing on a formula for the asymptotic behaviour of the degeneracy of dyons, first proposed in \cite{Dijkgraaf:1996it}, we find a series of subleading Bessels of the form
\begin{equation}\label{intro inst bessels}
 d(q)\simeq \sum_{\mu=1}^{P^2/2}\;\sum_{\substack{m\geq 0\\ \tilde{\Delta}_{\mu,m}>0}}\,\tilde{c}_{\mu}(P^2,m)\,I_{k+3/2}\left(\pi \sqrt{Q^2\tilde{\Delta}_{\mu,m}(P^2)}\right)+\mathcal{O}(e^{\pi\sqrt{Q^28n_p}})
\end{equation}
 with 
\begin{equation}
 \tilde{\Delta}_{\mu,m}(P^2)=8\left(n_p-m+\frac{\mu^2}{2P^2}\right)
\end{equation}
and $\tilde{c}_{\mu}(P^2,m)$ are some constant coefficients.
For $Q^2\gg 1$ and fixed $P^2$, the leading term in this sum, that is, the term with $m=0$ and $\mu=P^2/2$, reproduces the Bessel (\ref{Bessel n=4}), whereas the subleading terms are suggestive of a Rademacher type of expansion \cite{Dabholkar:2012nd}. In contrast, for the $\CN=8$ case these corrections are not present, with (\ref{Bessel n=8}) being corrected at a much subleading order by terms of the form $\exp{[\pi \sqrt{Q^2P^2}/n]}$, with $n$ an integer greater than one. These corrections are also present in the $\CN=4$ case but they will not be an object of study.

In this work, the focus is on the exact computation of the Bessel functions (\ref{Bessel n=8}) and (\ref{Bessel n=4}) including the precise coefficients. The emphasis will be on the one-quarter BPS case but we review at the same time the case of one-eighth BPS states, in particular, the measure proposed in \cite{Dabholkar:2011ec}.  Furthermore, we argue that the subleading corrections in (\ref{intro inst bessels}) may originate from instanton contributions to the $\text{AdS}_2$ path integral. Under certain assumptions, we determine not only the exact Bessel functions but we also compute an instanton corrected measure, which reproduces the exact coefficients $\tilde{c}_{\mu}(P^2,m)$ in the microscopic answers (\ref{intro inst bessels}).

The derivation for both one-eighth and one-quarter BPS black holes is similar. This follows from a $\CN=4$ truncation of $\CN=8$ supergravity, which allows us to see the one-eighth BPS black hole effectively as a one-quarter BPS black hole of $\CN=4$ supergravity. The truncation consists in projecting onto $(-1)^{F_L}$ even states in which we set the RR and RNS fields to zero, with $F_L$ the world-sheet fermion number \footnote{The $\mathcal{N}=8$ subalgebra splits into a left and right $\mathcal{N}=4$ subalgebras, respectively $(-1)^{F_R}$ and $(-1)^{F_L}$ invariant. The truncation consists in keeping the fields that transform under the right $\CN=4$ subalgebra.}. On the microscopic side, this is also consistent as the counting is valid only for charges vectors that are purely NSNS or that can be brought to such a configuration by a duality transformation \cite{Sen:2008sp}. Equivalently the counting is performed in a region of the moduli space invariant under a right $\CN=4$ subalgebra, the one that is $(-1)^{F_L}$ invariant; the one-eighth BPS states are effectively one-quarter BPS states of that subalgebra. For this reason, we can treat both supersymmetric examples in a similar way with the difference that in the one-eighth BPS case we need to take into account the contribution from the odd fields, that is, the fields which are odd under $(-1)^{F_L}$. 

To carry out this task, our starting point is a formula for $Z_{AdS_2}$ that one obtains using localization in supergravity \cite{Banerjee:2009af,Dabholkar:2010uh}. It was argued in \cite{Dabholkar:2010uh} that the path integral of $\CN=2$ off-shell supergravity on $AdS_2\times S^2$ reduces to the finite dimensional integral 
\begin{equation}\label{loc integral intro}
 Z_{AdS_2}\sim\int \prod^{n_V}_{a=0}d\phi^a \exp{\left[-\pi q_a\phi^a+\frac{\pi}{2}\text{Im}F(\phi+ip)\right]}
\end{equation}
where $(q,p)$ are respectively the electric and magnetic charges of the black hole. Here $F(X)$ is the $\mathcal{N}=2$ prepotential and $n_V$ is the number of $\CN=2$ vector-multiplets. The variables $\phi^a$ parametrize  a normalizable mode of the scalars that is left unfixed by the localization regulator. In a saddle point approximation of (\ref{loc integral intro}) one obtains precisely the area formula for the entropy.  

Despite this success, formula (\ref{loc integral intro}) lacks the correct measure that reproduces the microscopic answers (\ref{Bessel n=8}) and (\ref{Bessel n=4}). In a way, this is partially understood because the localization computation of \cite{Dabholkar:2010uh} only takes into account the vector-multiplets. Instead, the full answer must take into account not only the contribution from the other matter multiplets but also the contribution from the gravity multiplet, which in this context can be problematic. That is, if we want to apply localization in supergravity we need to deal with local supersymmetry and thus with a proper gauge fixed path integral, which is by itself a very difficult problem. Furthermore, the localization technique relies on a certain cohomological structure of the underlying space- equivariant cohomology to be more precise, and hence it is not clear how this can be translated to gravity, which can raise questions of background independence. Despite these issues, the problem of computing the localization one-loop determinants for the matter multiplets has been addressed recently in \cite{Murthy:2015yfa,Gupta:2015gga} using $\CN=2$ off-shell supergravity on $AdS_2\times S^2$. Here we give a qualitatively different approach based mostly 
on three dimensional supergravity but it includes the contribution from all the massless fields. In particular, we argue that the Kahler dependence of the one-loop determinants and the induced measure claimed in \cite{Murthy:2015yfa,Gupta:2015gga} for the $\CN=8$ and $\CN=4$ examples do not hold. We explain, nevertheless, why the approach for $\CN=8$ followed in \cite{Murthy:2015yfa} correctly reproduces the exact microscopic answer and how it can be correctly extended for the $\CN=4$ case.

To achieve our goal, we use well-known properties of $\text{AdS}_2$ and its relation to $\text{AdS}_3$ holography to derive the measure for the integral (\ref{loc integral intro}). Partially justified by the computation of \cite{Gupta:2012cy} which fixes the background to be exactly $AdS_2\times S^2$, our derivation is based on the assumption that the full $AdS_2$ path integral has the form (\ref{loc integral intro}), that is, the contribution from other multiplets enters only through the measure.  After all, the integral (\ref{loc integral intro}) reproduces correctly the area formula in all known examples and so it is not expected that other fields contribute already at the exponential level. 

The key idea is as follows. We map the problem of computing the measure to a certain anomaly in the path integral of super Chern-Simons theory on $\text{AdS}_3$. From a bulk point of view, this anomaly is related to a dependence on a metric choice for the Chern-Simons path integral. Using holography, we can relate this dependence to the modular weight of the dual $\text{CFT}_2$ partition function- for the supersymmetric black holes this partition function is an elliptic genus. It is well known that in general the $\text{CFT}_2$ partition function on the torus is not invariant under global diffeomorphisms but transforms covariantly with a certain weight under modular transformations. Equivalently, we can say that the partition function is anomalous under $SL(2,\mathbb{Z})$ diffeomorphisms. In turn, the low/high temperature modular transformation can be used to show that the asymptotic growth of the Fourier coefficients of the $\text{CFT}_2$ partition function \cite{Dijkgraaf:2000fq}, and thus the $\text{AdS}_2$ partition function, are of Bessel type. Therefore, by understanding the anomalous modular transformation in the bulk we have immediate access to the structure of perturbative corrections to the entropy, which are determined by the leading Bessel in (\ref{Bessel n=4}).  Nevertheless, this is not the full answer. In going from the gravity picture to the Chern-Simons formulation we need to keep track of an anomalous field redefinition- it is anomalous due to zero modes. This contribution, in turn, can be identified with a certain degeneracy that accompanies the Bessel function, which, in the theory of Jacobi forms, can be identified with a ``polar'' coefficient.

To compute this anomaly we use three dimensional supergravity. By convenience, this can be written as a super Chern-Simons action based on the gauge group $SU(1,1|2)_R\times SU(1,1)_L\times SU(2)_L$. On top of this, we will consider additional abelian Chern-Simons terms. We put this theory on $AdS_2\times S^1$, which is the same as the quotient of global $AdS_3$ by an additive group $\Gamma$, and use microcanonical boundary conditions consistent with the $AdS_2$ path integral. We explore the orbifold construction to argue that the one-loop contribution in Chern-Simons theory must hold for any value of the charges. After all, this leads to a correction of the form $\ln |\Gamma|$ to the effective action, with $|\Gamma|$ the order of the group, and therefore cannot be renormalized by a local counterterm \footnote{By this we mean that any local counterterm evaluated on the on-shell solution will be a polynomial in $|\Gamma|$ due to the orbifold. As such, no logarithmic correction can be produced.}. In particular we show that the one-loop approximation to the $\text{AdS}_2$ partition function has the form
\begin{equation}
Z_{\text{AdS}_2}|_{\text{1-loop}}=\sum_{A}\exp{[\text{CS}(A,M)]}\,Z_{\text{1-loop}}^{\text{CS}}(|\Gamma|)
\end{equation}
where $A$ is a flat connection and $\text{CS}(A,M)$ is the Chern-Simons action  on $M=\text{AdS}_3/\Gamma$ properly renormalized.  In particular, we find that the one-loop correction is
\begin{equation}
Z_{\text{1-loop}}^{\text{CS}}(|\Gamma|)=\vartheta\,\frac{|\Gamma|}{\sqrt{\tilde{k}_Lk_L}}\left(\frac{|\Gamma|}{p^1}\right)^{N_V/2}
\end{equation}
where $\vartheta$ is the size of $AdS_2\times S^1$ in the physical theory, $\tilde{k}_L$ and $k_L$ are respectively the $SU(1,1)_L$ and $SU(2)_L$  levels of the non-supersymmetric Chern-Simons terms, and $p^1$ is the abelian Chern-Simons level. By identifying the parameter $|\Gamma|$ with the variable $\phi^0$ in the integral (\ref{loc integral intro}), we argue that the component $Z_{\text{1-loop}}^{\text{CS}}(|\Gamma|)$ uniquely determines  the measure in the integral (\ref{loc integral intro}). 

%Their contribution can be fixed using a certain scaling symmetry of two derivative supergravity explored also in \cite{Sen:2009bm}. Under this symmetry the dilaton gets shifted by $\ln\lambda$, the NSNS fields remain unchanged and the RR fields are scaled by $\lambda$.

In essence, the main result is an exact formula for $Z_{AdS_2}$, which in the $\CN=8$ case takes the form 
  
\begin{equation}\label{main}
 Z_{AdS_2}^{\CN=8}=\int \prod_{a=0}^{N_V+1}d\phi^a \mathcal{M}_{1/8}(\phi^0)\exp{\left[-\pi q_a\phi^a+\frac{\pi}{2}\text{Im}F(\phi+ip)\right]}\times Z_{\text{odd}}
\end{equation}with 
\begin{equation}\label{measure intro}
 \mathcal{M}_{1/8}(\phi^0)=\frac{P^2}{\phi^0p^1},\;\;\;Z_{\text{odd}}=(P^2)^{-4}
\end{equation}
where $F(X^I)=X^1C_{ab}X^aX^b/X^0$ is the tree level prepotential, $Z_{\text{odd}}$ is the contribution from the odd fields and $N_V$ is the number of vector-multiplets of the $\CN=4$ truncation. This formula correctly reproduces the Bessel answer (\ref{Bessel n=8}). Furthermore, it gives a fundamental principles derivation of the measure proposed in \cite{Dabholkar:2011ec}, as we wanted to show.

In the $\CN=4$ case, the partition function has a similar expression except that the contribution from the odd fields is absent and the prepotential $F(X)$ is modified by instanton corrections. We compute the exact zero instanton $\text{AdS}_2$ path integral,
 \begin{equation}\label{main2}
 Z_{AdS_2}^{\CN=4}= \int\prod_{a=0}^{N_V+1}d\phi^a\, \mathcal{M}_{1/4}(\phi^0)\,\exp{\left[-\pi q_a\phi^a+\frac{\pi}{2}\text{Im}F^{(0)}(\phi+ip)\right]}
\end{equation}
with 
\begin{equation}\label{intro measure 1/4}
 \mathcal{M}_{1/4}(\phi^0)=\frac{P^2+4c_1}{\phi^0p^1}
\end{equation}
Here $F^{(0)}(X)$ is the zero instanton prepotential $F^{(0)}=X^1C_{ab}X^aX^b/X^0+c_1\,\hat{A} X^1/X^0$, with $c_1$ taking  values $1,0$ for the $K3$ and $T^4$ CHL models respectively, and $N_V$ is the number of $\CN=4$ vectormultiplets. Note that for large $P^2$ the zero instanton measure (\ref{intro measure 1/4}) reproduces the one-eighth BPS measure (\ref{measure intro}). This is as expected since the measure $\mathcal{M}(\phi)$ comes entirely from the $\CN=4$ supergravity multiplet and therefore for large charges it should become the same in all examples. 

The formula (\ref{main2}) leads precisely to the microscopic answer (\ref{Bessel n=4}), including the precise coefficients that multiply the Bessel function. Furthermore, we argue that corrections due to instantons can be incorporated by integrals similar to (\ref{main2}) and we compute the exact measure in agreement with microscopics. The idea relies on the observation that the integral (\ref{loc integral intro}) with the instanton corrected prepotential suggests a similar Chern-Simons computation but with renormalized levels. From this, it follows an unitarity condition that truncates the instanton sum and leads to the same tail of Bessel functions (\ref{intro inst bessels}) that we find from microscopics. 

The plan of the paper is as follows. In section \S\ref{micro sec} we study the exact microscopic answers for both one-eighth and one-quarter BPS states. We derive the exact Bessel function that captures all perturbative corrections to the area formula and we discuss the role of the subleading contributions. In section \S\ref{localization sec} we first describe the on-shell $\CN=4$ background and then we review the application of localization techniques in the computation of the $\text{AdS}_2$ path integral. Finally, in section \S\ref{measure sec} we describe the computation of the measure. We divide this section into three main parts. First, we consider the one-eighth BPS case and the contribution coming from the odd fields and then we determine both the $\CN=4$ vector and supergravity  multiplet measures using the Chern-Simons formulation. We conclude with a discussion about open problems and other future directions.

 \section{Microscopic degeneracy }\label{micro sec}

In the following sections we describe the microscopic partition functions that capture the spectrum of one-eighth and one-quarter BPS states. We are mainly interested in the behaviour of the degeneracy for sufficiently large charges. For the $\CN=8$ case we review a formula derived originally in \cite{Dabholkar:2011ec}. In the $\CN=4$ case, we present a novel way to compute the leading Bessel function and subleading corrections by developing on the formula for the asymptotic degeneracy of dyons \cite{Dijkgraaf:1996it,David:2006yn}.

\subsection{$\mathcal{N}=8$ string theory and one-eighth BPS states}\label{micro N=8 sec}

We consider IIB string theory compactified on $T^4\times \tilde{S}^1\times S^1$. An important difference with the $\CN=4$ case is that here we consider only a subspace of all the one-eighth BPS configurations. In particular, we consider those that carry only NS-NS charges  or that can be mapped to this case after an U-duality transformation \cite{Sen:2008ta}. The microscopic formula that we present is valid in a region of moduli space where the RR moduli are turned off. Effectively we are counting one-quarter BPS states of a right $\CN=4$ subalgebra of the $\CN=8$ supersymmetry algebra \footnote{The region of moduli space where the counting is done is invariant under a left and right $\CN=4$ subalgebras. }. In this region of the moduli space, the U-duality group is broken to $SL(2,\mathbb{Z})\times SO(6,6;\mathbb{Z})$.

 The one-eighth BPS configuration that was considered in \cite{Maldacena:1999bp,Sen:2008sp} consists of a D5-brane wrapped on $T^4\times S^1$, $Q_1$ D1-branes wrapped along $S^1$, a Kaluza-Klein monopole associated with the circle $\tilde{S}^1$, $n$ units of momentum along the circle $S^1$ and $J$ units of momentum along the circle $\tilde{S}^1$. This configuration can be mapped to a purely NS-NS configuration in IIA after a set of U-duality transformations as described in \cite{Sen:2008ta}.

The index that captures one-eighth BPS states has the expression
\begin{equation}
 d(Q,P)=(-1)^{Q.P+1}c(\Delta), 
\end{equation}
with $c(u)$ the Fourier coefficients of the Jacobi form \footnote{In general the Fourier coefficient $c(u)$ of a Jacobi form with index $k$ depends not only on $4kn-l^2$ but also on $l\,\text{mod}(2k)$. In this case we have $k=1$ and thus $4n-l^2$ is even or odd when $l$ is too.}
\begin{equation}
 -\frac{\vartheta(\tau,z)^2}{\eta(\tau)^6}=\sum_{k,l}c(4k-l^2)e^{2\pi i(k\tau+lz)}.
\end{equation}
Here $\Delta=Q^2P^2-(Q.P)^2$ with $Q^2$, $P^2$ and $Q.P$ are the $SO(6,6;\mathbb{Z})$ T-duality invariants. 

The coefficients $c(u)$ admit an exact Rademacher expansion \cite{Dabholkar:2011ec}. This is an exact formula for the Fourier coefficients of Jacobi forms of non-positive weight. In essence, it consists of an infinite but convergent sum of Bessel functions. In this case, the leading behaviour is controlled by
\begin{equation}\label{deg bessel 1/8}
 c(\Delta)=2\pi (\frac{\pi}{2})^{7/2}\frac{1}{2\pi i}\int_{\epsilon -i\infty}^{\epsilon +i\infty}\frac{dt}{t^{9/2}}e^{t+\frac{\pi^2\Delta}{4t}}+\ldots
\end{equation}
where the $\dots$ refer to terms that are exponentially supressed. Furthermore,  it is convenient to write the integral in form 
\begin{equation}\label{deg 1/8 integral}
 c(\Delta)=\frac{1}{\sqrt{2}i\pi}\int_{\mathcal{C}}\frac{d\tau_1d\tau_2}{(\tau_2)^6}\frac{e^{-K}}{(P^2)^{4}}\exp{\left[\frac{\pi}{2\tau_2}|Q+\tau P|^2\right]}+\ldots
\end{equation}
\footnote{In the modulus square in (\ref{deg 1/8 integral}), there is an implicit contraction with the T-duality invariant tensor.} where we defined
\begin{equation}
 e^{-K}\equiv\tau_2\pi P^2.
\end{equation}
 The contour $\mathcal{C}$ takes $\tau_1$ over the imaginary axis and $\tau_2$ over the axis $\epsilon+i\mathbb{R}$ with $\epsilon>0$. This form of the integral will be useful later on, as a way to physically check the bulk computation of the $\CN=4$ answer.

% The exponentially suppressed terms have a similar Bessel function behavior with coefficients given by $SL(2,\mathbb{Z})$ Kloosterman sums \cite{Dabholkar:2011ec}, that is,
% \begin{equation}\label{kloos1}
%  \propto \sum_{c>1}^{\infty}c^{-9/2}Kl_c(\Delta)\int_{\epsilon -i\infty}^{\epsilon +i\infty}\frac{dt}{t^{9/2}}e^{t+\frac{\pi^2\Delta}{4tc^2}}
% \end{equation}
% where 
% \begin{equation}\label{kloos2}
%  Kl_{c}(\D) := \sum_{\genfrac{}{}{0pt}{}{-c \leq d< 0;}{(d,c)=1}} e^{2\pi i \frac{d}{c} (\D/4)} \; M^{-1}(\gamma_{c,d})_{\nu 1} \; 
% e^{2\pi i \frac{a}{c} (-1/4)} 
% \end{equation}
% with $\nu=\Delta \text{ mod}\,2$ and $ad=1\text{ mod}c$. 

\subsection{$\mathcal{N}=4$ string theory and one-quarter BPS states}\label{micro N=4 sec}

In this section we consider one-quarter BPS states in $\CN=4$ CHL compactifications. These are particular $\mathbb{Z}_N$ orbifolds of IIB string theory on either $K3\times T^2$ or $T^4\times T^2$.  We present a summarized description of the spectrum and BPS-state counting for both cases.  The discussion presented is completely systematic in $N$. This is advantageous for a comparison with the gravitational computation and it will help us highlight the key points of our derivation. For a more detailed description of CHL compactifications and BPS-state counting we point the reader to \cite{Sen:2007qy} and  references therein. 

One of the goals of this section is to rewrite the microscopic degeneracies in a manner suitable for a gravitational comparison. We will find that the microscopic answer can be written as a finite sum of Bessel functions, up to much subleading terms. 
 
We consider IIB string theory on $M\times \tilde{S}^1\times S^1$, with $M=K3,\,T^4$ modded out by a $\mathbb{Z}_N$ symmetry group. The orbifold identification involves a $1/N$ shift along the circle $S^1$ and an order $N$ $\tilde{g}$ transformation on $M$. The element $\tilde{g}$ commutes with the $\mathcal{N}=4$ supersymmetry generators and therefore the orbifold preserves all the supersymmetry of the parent theory. By convention, we take the radius of the circle $S^1$ to have size $N$ in the parent theory. Here $N$ runs over $1,2,3,5,7$ in the $K3$ case and $2,3$ in the $T^4$ case. Under the orbifold only a subgroup $SL(2,\mathbb{Z})\times SO(6,r-6;\mathbb{Z})$ of the U-duality group survives, with $r=2k+8$ the total number of $U(1)$ gauge fields. The integer $k$ is given by
\begin{eqnarray}\label{N and k}
 &&k=\frac{24}{N+1}-2,\;\;N=1,2,3,5,7,\text{ for }M=K3\\
&&k=\frac{12}{N+1}-2,\;\;N=2,3\text{ for }M=T^4.
\end{eqnarray}

Let us consider a configuration with a single D5-brane wrapping $M\times S^1$, $Q_1$ D1-branes wrapping $S^1$, a single Kaluza-Klein monopole associated with the circle $\tilde{S}^1$,  $n/N$ units of momentum along $S^1$ and $J$ units of momentum along $\tilde{S}^1$. At low energies this system is described by a two dimensional $(0,4)$ SCFT on $\mathbb{R}\times S^1$. At the orbifold point in the moduli space, this theory is described by a symmetric product sigma model and we can compute the supersymmetric index that counts one-quarter BPS states. The index, which we denote by $d(Q,P)$, has the form
\begin{equation}\label{deg 1/4}
 d(Q,P)=(-1)^{Q.P+1}\frac{1}{N}\int_{\mathcal{C}}d\rho d\sigma dv \,\frac{e^{-\pi i(N Q^2\rho +P^2/N \sigma +2Q.P v)}}{\Phi_N(\rho,\sigma,v)}
\end{equation}where $\mathcal{C}$ is a three dimensional contour in the complexified $(\rho,\sigma,v)=(\rho_1+iI_1,\sigma_1+iI_2,v_1+iI_3)$ space with
\begin{eqnarray}
 &&I_1,I_2,I_3=\text{constant}\gg1 \\
&&0\leq\rho_1\leq 1,\;0\leq\sigma_1\leq N,\;0\leq v_1\leq 1.
\end{eqnarray}
and $Q^2=2n/N$, $P^2=2Q_1$ and $Q.P=J$ are the T-duality invariants. The function $\Phi_N(\rho,\sigma,v)$ is a $Sp(2,\mathbb{Z})$ modular form. In particular, for $N=1$ it  becomes the Igusa cusp form: the unique weight ten Siegel modular form, while for other $N$ we obtain Siegel modular forms of congruence subgroups. 
 
We now study the degeneracy in the regime of large charges. In a saddle point approximation of the integral (\ref{deg 1/4}) we deform the contour and pick poles of $1/\Phi_N(\rho,\sigma,v)$. The leading contribution comes from the residue at a quadratic divisor of $\Phi_N(\rho,\sigma,v)$ and has final expression  \cite{Dijkgraaf:1996it,Shih:2005he,David:2006yn, Sen:2007qy},
\begin{eqnarray}\label{deg 1/4 integral}
 d(Q,P)\simeq&&\frac{(-1)^{Q.P}}{4\pi N^{(k+4)/2}}\int \frac{d\tau_1d\tau_2}{\tau_2^2}\left[2(k+3)+\frac{\pi}{\tau_2}|Q-\tau P|^2\right]\times\\ \nonumber
&&\exp{\left[\frac{\pi}{2\tau_2}|Q-\tau P|^2-\ln g(\tau)-\ln g(-\bar{\tau})-(k+2)\ln(2\tau_2)\right]}+\ldots
\end{eqnarray}
with $\tau=\tau_1+i\tau_2$ and $\bar{\tau}=\tau_1-i\tau_2$. The function $g(\tau)$ is determined by the pole structure of $1/\Phi_N(\rho,\sigma,v)$. In the case of $M=K3$ it is given by
\begin{equation}\label{K3 g(t)}
 g(\tau)=\eta(\tau)^{k+2}\eta(N\tau)^{k+2},
\end{equation}while for $M=T^4$ it has the form
\begin{equation}\label{T4 g(t)}
 g(\tau)=\eta(\tau)^{\frac{2N(k+2)}{N-1}}\eta(N\tau)^{-2\frac{k+2}{N-1}},
\end{equation}
with $N,k$ given as in (\ref{N and k}). 

The formula (\ref{deg 1/4 integral}) is not in a form that is suitable for a comparison with the localization computation. The reason is that the measure in (\ref{deg 1/4 integral}) depends on the electric charges, while from a gravitational point of view they appear only at level of the Wilson lines. Besides, the contour in (\ref{deg 1/4 integral}) has to be chosen appropriately. The only requirement at this point is that it passes near the leading physical saddle \footnote{By physical, we mean that it reproduces the attractor background and thus the area formula for the entropy.}. We show that there is a choice for which we recover not only the leading Bessel function (\ref{Bessel n=4}) \cite{Dabholkar:2010rm} but also an additional tail of subleading Bessel type corrections. Namely, we choose a contour with  $\tau_1, \tau_2$ complex defined as
\begin{eqnarray}\label{contour 1/4}
 \hat{\mathcal{C}}:\;&&\tau_1=i\tau_2 u,\;\;-1+\delta\leq u\leq 1-\delta,\nonumber\\
&&\tau_2=\epsilon+iy,\;\;-\infty<y<\infty,\;\epsilon>0\nonumber\\
{}
\end{eqnarray}
Here $\delta$ is small but positive (we will make precise what we mean by small in due course). This contour ensures that we always have $\Im(\tau)$ and $\Im(\bar{\tau})$ positive- here $\tau$ and $\bar{\tau}$ are not necessarily complex conjugate. This in turn leads to the exact Bessel function determined in \cite{Dabholkar:2012nd} for the leading asymptotics of one-quarter BPS states in IIB on $K3\times T^2$.

We proceed with an integration by parts. First we rewrite the expression (\ref{deg 1/4 integral}) in the convenient way
\begin{equation}\label{entropyfnct2}
 d(Q,P)\simeq (-1)^{Q.P+1}\int \frac{d^2\tau}{\tau_2^{k+4}}\left[2(k+3)+\pi\frac{|Q-\tau P|^2}{\tau_2}\right]e^{\frac{\pi}{2}\frac{|Q-\tau P|^2}{\tau_2}-\Omega(\tau,\bar{\tau})}
\end{equation}with
\begin{equation}
 \Omega(\tau,\bar{\tau})=\ln g(\tau)+\ln g(-\bar{\tau}).
\end{equation}
The exponential is just the entropy function of Sen (\ref{ent fnct sen})
\begin{equation}\label{entropyaction}
 \mathcal{E}=\frac{\pi}{2}\frac{|Q-\tau P|^2}{\tau_2}-\Omega(\tau,\bar{\tau}).
\end{equation}
Using the identity
\begin{equation}
 \frac{\partial}{\partial \tau_2}\frac{|Q-\tau P|^2}{\tau_2}=-\frac{|Q-\tau P|^2}{\tau_2^2}+2P^2
\end{equation} we can write the measure in (\ref{entropyfnct2}) as
\begin{equation}
 \frac{1}{\tau_2^{k+3}}\left[\frac{2(k+3)}{\tau_2}+\pi\frac{|Q-\tau P|^2}{\tau_2^2} \right]=\frac{1}{\tau_2^{k+3}}\left[\frac{2(k+3)}{\tau_2}-2\frac{\partial}{\partial \tau_2}\mathcal{E}-2\frac{\partial}{\partial \tau_2}\Omega+2\pi P^2\right]
\end{equation}
which leads to
\begin{equation}
 d(Q,P)\simeq (-1)^{Q.P+1}\int \frac{d^2\tau}{\tau_2^{k+3}}\left[\frac{2(k+3)}{\tau_2}-2\frac{\partial}{\partial \tau_2}\mathcal{E}-2\frac{\partial}{\partial \tau_2}\Omega+2\pi P^2\right]e^{\mathcal{E}}.
\end{equation}The first two terms on the R.H.S. can be written as a total derivative
\begin{equation}
 \int \frac{d^2\tau}{\tau_2^{k+3}}\left[\frac{2(k+3)}{\tau_2}-2\frac{\partial}{\partial \tau_2}\mathcal{E}\right]e^{\mathcal{E}}=-2\int d^2\tau\frac{\partial}{\partial \tau_2}\left(\frac{e^{\mathcal{E}}}{\tau_2^{k+3}}\right)
\end{equation}
which vanishes for the contour (\ref{contour 1/4}). Hence, the final expression for the degeneracy is
\begin{equation}\label{OSV 1/4}
 d(Q,P)\simeq 2(-1)^{Q.P+1}\int_{\hat{\mathcal{C}}} \frac{d^2\tau}{\tau_2^{k+4}}e^{-K}e^{\mathcal{E}}
\end{equation}with
\begin{equation}
 e^{-K}\equiv\tau_2\left[\pi P^2-\frac{\partial}{\partial \tau_2}\Omega\right].
\end{equation}

Note the similarities between the one-quarter  and the one-eighth BPS formulas, respectively (\ref{OSV 1/4}) and (\ref{deg 1/8 integral}). In particular, if we neglect the factor of $(P^2)^{-4}$ in the one-eighth BPS formula (\ref{deg 1/8 integral}), then both integrands have the form of the exponential of the entropy function $\mathcal{E}$ times the quantum corrected K\"{a}hler potential $e^{-K}$ \cite{Shih:2005he} and a factor of $\tau_2^{-r/2}$, where $r$ is the total number of $U(1)$ vector fields of the $\CN=4$ supergravity (truncation in the one-eigth BPS case).  

We can proceed further and expand the modular functions in $\Omega(\tau,\bar{\tau})$ as a Fourier series in powers of $q=\exp{(2\pi i\tau)}$  and $\bar{q}=\exp{(-2\pi i\bar{\tau})}$. Note that we have always $|q|<1$ for the contour (\ref{contour 1/4}) and thus we expand as 
\begin{eqnarray}\label{fourier expansion}
 \exp{(-\Omega(\tau,\bar{\tau}))}&=&\left(\sum_{n=0}^{\infty}d(n)q^{n-n_p}\right)\left(\sum_{m=0}^{\infty}d(m)\bar{q}^{m-n_p}\right)\nonumber\\
&=&|q|^{-2n_p}\sum_{n'=0}^{\infty}e^{-2\pi n' \tau_2}\sum_{m'=0}^{n'}d(n'-m')d(m')e^{2\pi i(n'-2m')\tau_1}
\end{eqnarray}
with $d(n)$ the Fourier coefficients 
\begin{equation}\label{g Fourier coeff}
 g(\tau)^{-1}=q^{-n_p}\sum_{n=0}^{\infty}d(n)q^{n}.
\end{equation}
Here $n_p=0,1$ for $T^4$ and $K3$ orbifolds respectively.

Using the expansion (\ref{fourier expansion}) we can write formula (\ref{OSV 1/4}) as the sum
\begin{eqnarray}\label{CHL micro inst OSV}
 d(Q,P)\simeq  &&(-1)^{Q.P+1}\sum_{n=0}^{\infty}\pi\Big(P^2+4n_p-2n\Big)\sum_{m=0}^{n}d(n-m)d(m)\times \nonumber\\
&&\times \int_{\hat{\mathcal{C}}} \frac{d^2\tau}{\tau_2^{k+3}}\exp{\left[\frac{\pi}{2}\frac{|Q-\tau P|^2}{\tau_2}+\pi (4n_p-2n)\tau_2+2\pi i(n-2m)\tau_1\right]}.
\end{eqnarray}
% \begin{eqnarray}\label{inst exponential2}
% \frac{\pi}{2}\frac{|Q-\tau P|^2}{\tau_2}+\pi\tau_2(4n_p-2n)+2\pi i\tau_1(n-2m)=\nonumber\\
% \frac{\pi}{2}\frac{\Delta/P^2}{\tau_2}+\frac{\pi}{2}\tau_2\left(P^2+8n_p-4n +4(n-2m)^2/P^2\right)+\nonumber\\
% \frac{\pi}{2}\frac{P^2}{\tau_2}\left(\tau_1-\frac{Q.P}{P^2}+2i\frac{\tau_2}{P^2}(n-2m)\right)^2+2\pi i\frac{Q.P}{P^2}(n-2m)
% \end{eqnarray} 
We can massage further the exponential and write the degeneracy in the form
\begin{equation}\label{sum over J}
 d(Q,P)\simeq  (-1)^{Q.P+1}\sum_{n=0}^{\infty}\pi\Big(P^2+4n_p-2n\Big)\sum_{m=0}^{n}d(n-m)d(m)e^{2\pi i\frac{Q.P}{P^2}(n-2m)}\,\mathcal{J}_{(n,m)}(Q,P)
\end{equation}
with the integral $\mathcal{J}_{(n,m)}$ defined as
  \begin{eqnarray}\label{J integral}
 \mathcal{J}_{(n,m)}=&&\int_{\hat{\mathcal{C}}} \frac{d\tau_2}{\tau_2^{k+3}}\exp{\left[\frac{\pi}{2}\frac{\Delta/P^2}{\tau_2}+4\pi\tau_2\mathcal{F}(n,m;P^2)\right]}\times\nonumber\\
&&\times\int d\tau_1\exp{\left[\frac{\pi}{2}\frac{P^2}{\tau_2}\left(\tau_1-\frac{Q.P}{P^2}+2i\frac{\tau_2}{P^2}(n-2m)\right)^2\right]}
\end{eqnarray}
with
\begin{equation}\label{Besse arg}
\mathcal{F}(n,m;P^2)\equiv n_p-m +\frac{l(n,m)^2}{2P^2},
\end{equation}
and
\begin{equation}
 l(n,m)=P^2/2-(n-2m).
\end{equation}

Lets focus on the $\tau_1$ integral for the moment. After the change of variables $\tau_1=i\tau_2 u$ it becomes
\begin{equation}\label{u integral}
 \mathcal{I}_{u}=i\tau_2\int_{-1+\delta}^{1-\delta} du\exp{\left[-\frac{\pi}{2}P^2\tau_2\left(u+r-\frac{Q.P}{i\tau_2P^2}\right)^2\right]}
\end{equation}
where we have defined $r\equiv 2(n-2m)/P^2$. For large values of $|\tau_2|$, we can perform the $u$ integral by a saddle point approximation. In this case, given the contour $\hat{\mathcal{C}}$, it is enough to take $\Re(\tau_2)=\epsilon>>1$. Moreover, in this limit we can neglect the term $Q.P/i\tau_2 P^2$ in the square and thus we can write the integral (\ref{u integral}) as
\begin{equation}\label{u integral 2}
 \mathcal{I}_u\simeq i\sqrt{\frac{2\tau_2}{\pi P^2}}\int^{\sqrt{\pi P^2\tau_2/2}(1+r-\delta)}_{\sqrt{\pi P^2\tau_2/2}(-1+r+\delta)} dz\, e^{-z^2}
\end{equation}
In computing this integral by saddle point approximation, there are two cases to consider. The most relevant is when
\begin{equation}
 |r|\leq 1-\delta.
\end{equation}
In this case, the saddle is inside the contour of integration and thus the saddle point approximation of (\ref{u integral 2}) is simply
\begin{equation}
 \mathcal{I}_u\simeq i\sqrt{\frac{2\tau_2}{\pi P^2}}\int_{-\infty}^{\infty}dz\, e^{-z^2}\simeq i\sqrt{\frac{2\tau_2}{P^2}},\qquad |r|\leq 1-\delta
\end{equation}
up to terms that are exponentially decaying. In the other case, that is, when $|r|>1-\delta$ we can use the asymptotics of the complemetanty error function 
\begin{equation}\label{error function}
 \int_{a}^{\infty} dx\,e^{-x^2}\simeq \frac{2}{\pi a}e^{-a^2}+\mathcal{O}(e^{-a^2}/a^2),\;a\gg1
\end{equation}
to estimate
\begin{equation}
 \mathcal{I}_{u}\simeq i\frac{4}{\pi^2 P^2}\frac{e^{-\frac{\pi P^2\tau_2}{2}(|r|-1+\delta)^2}}{(|r|-1+\delta)},\qquad |r|>1-\delta
\end{equation}
Putting these results back in the integral (\ref{J integral}) we find two types of asymptotic behaviour. In the first,  when $|r|\leq 1-\delta$, we obtain
\begin{equation}\label{J r<1}
 \mathcal{J}_{(n,m)}\simeq i\sqrt{\frac{2}{P^2}}\int_{\epsilon-i\infty}^{\epsilon+i\infty} \frac{d\tau_2}{\tau_2^{k+5/2}}\exp{\left[\frac{\pi}{2}\frac{\Delta/P^2}{\tau_2}+4\pi\tau_2\mathcal{F}(n,m;P^2)\right]}
\end{equation}
which is of Bessel type. However, if $\mathcal{F}(n,m;P^2)\leq 0$ we can close the contour at infinity and since there is no pole inside we obtain zero. In turn, this leads to the condition 
\begin{equation}\label{positivity cond}
 \mathcal{F}(n,m;P^2)>0,\qquad |r|\leq 1-\delta
\end{equation}
At this point it is convenient to impose the condition that $\delta P^2/2<1$ such that the formula (\ref{J r<1}) is valid exactly for $|r|<1$. We assume this for now on.

In the case when $|r|\geq 1$ the asymptotics are governed instead by
\begin{eqnarray}
 &&\mathcal{J}_{(n,m)}\simeq  i\frac{4}{\pi^2 P^2(|r|-1+\delta)}\times\nonumber \\
&&\int_{\epsilon-i\infty}^{\epsilon+i\infty}\frac{d\tau_2}{\tau_2^{k+3}}\exp{\left[\frac{\pi}{2}\frac{\Delta/P^2}{\tau_2}+4\pi\tau_2\left(\mathcal{F}(n,m;P^2)-\frac{P^2}{8}(|r|-1+\delta)^2\right)\right]}
\end{eqnarray}
which is still of Bessel type but has different index. By the same argument that gives the condition (\ref{positivity cond}), this integral will be non-zero only when the term proportional to $\tau_2$ in the exponential is positive. In this case we have to truncate further to the terms with $m=0$ for $r>0$, and $n-m=0$ for $r<0$, with the condition that $1\leq |r|<|r^*|$, for a maximum $r^*$ that solves the equation $n_p-\delta P^2(|r^*|-1)/4-\delta^2P^2/8=0$. Note that, when $n_p=0$, that is, for $T^4$ orbifolds, the exponential is always negative and thus these terms are not present. Under these conditions we find
\begin{eqnarray}
 &&\mathcal{J}_{(n,m)}\simeq  i\frac{4}{\pi^2 P^2(|r|-1+\delta)}\times\nonumber \\
&&\int_{\epsilon-i\infty}^{\epsilon+i\infty}\frac{d\tau_2}{\tau_2^{k+3}}\exp{\left[\frac{\pi}{2}\frac{\Delta/P^2}{\tau_2}+4\pi\tau_2\left(n_p-\delta P^2(|r|-1)/4-\delta^2P^2/8\right)\right]},\qquad 1\leq |r|<|r^*|\nonumber\\
{}
\end{eqnarray}

With this analysis we conclude that the integral $\mathcal{J}(n,m)$ has two kinds of behaviour. For $|r|<1$ it behaves as a modified Bessel function with index $k+3/2$ while for $|r|\geq 1$ and $n_p= 1$ the  Bessel has index $k+2$. As we will see later, the ones which are of interest for us are the Bessels with index $k+3/2$. Moreover, these are the ones that do not depend on the regularization, that is, on the choice of $\delta$ for $\delta < 2/P^2$.

Given the conditions that $\mathcal{F}(n,m;P^2)>0$  and $-P^2/2< n-2m< P^2/2$ it is easy to show that $n$ is bounded by
\begin{equation}
  n< k+2n_p
\end{equation}
and thus the sum over the terms with $-P^2/2< n-2m< P^2/2$ is finite. This leads to an answer for the degeneracy which is a finite sum of Bessel functions, that is,
\begin{eqnarray}\label{finite sum Bessels}
 &&d(Q,P)\simeq (-1)^{Q.P+1}\sum_{n=0}^{P^2/2+2n_p-1}i\pi\Big(P^2+4n_p-2n\Big)\times\nonumber\\
&&\times\sum_{\substack{m\geq 0\\ 0\leq n-2m<P^2/2\\ \mathcal{F}(n,m,P^2)>0}}^{n}d(n-m)d(m)\left[2\cos(2\pi (n-2m)Q.P/P^2)-\delta_{n,2m}\right]\,\times\nonumber\\
&&\times \sqrt{\frac{2}{P^2}}\int_{\epsilon-i\infty}^{\epsilon+i\infty} \frac{d\tau_2}{\tau_2^{k+5/2}}\exp{\left[\frac{\pi}{2}\frac{\Delta/P^2}{\tau_2}+4\pi\tau_2\left(n_p-m +\frac{l(n,m)^2}{2P^2}\right)\right]}+\mathcal{O}(e^{2\pi\sqrt{\Delta n_p/k}})\nonumber\\
{}
\end{eqnarray}
with $\delta_{j,l}$ the Kronecker delta function. The terms of order $e^{2\pi\sqrt{\Delta n_p/k}}$ are Bessels of index $k+2$. Some of these can compete asymptotically with the other Bessels but since they depend explicitly on $\delta$, parameter for which we have some freedom to choose, we assume that they are not relevant for the physics we want to study.

In this work we are mainly interested in the zero instanton term which is the leading term in the tail (\ref{finite sum Bessels}). We find
\begin{eqnarray}\label{Bessels 1/4}
 d(Q,P)_{(n,m)=0}&\simeq&\frac{(P^2+4n_p)}{\sqrt{P^2}}\int_{\epsilon-i\infty}^{\epsilon+i\infty}\frac{dt}{t^{k+3-1/2}}\exp{\left[\frac{\pi^2\Delta}{4 t P^2}+t(P^2+8n_p)\right]}\nonumber\\
 &=&\frac{(P^2+4n_p)}{\sqrt{P^2}}(P^2+8n_p)^{k+3/2}\;I_{k+3/2}\left(\pi \sqrt{\Delta (1+8n_p/P^2)}\right).
\end{eqnarray}
 In particular, the two following examples are instructive. In the case of $M=K3$ and $N=1$ we obtain the Bessel function
\begin{equation}
 d(Q,P)_{(n,m)=0}\propto (P^2)^{-12}I_{23/2}(\pi\sqrt{\Delta})
\end{equation}
where we have taken $P^2$ large. This in perfect agreement with the results in \cite{Dabholkar:2010rm,Dabholkar:2012nd}. Also, in the case of $M=T^4$ and $N=2$  we obtain, in the same charge limit,
\begin{equation}
 d(Q,P)_{(n,m)=0}\propto (P^2)^{4}I_{7/2}(\pi\sqrt{\Delta}).
\end{equation}
Up to a factor of $(P^2)^{4}$, this is precisely the same Bessel function one obtains from the one-eighth BPS formula (\ref{deg 1/8 integral}). This will be important to understand the role of the odd fields in a $\CN=4$ truncation of $\CN=8$ supergravity.

\section{Black hole entropy and supersymmetric localization}\label{localization sec}

In the first part of this section we review the quantum entropy formalism introduced by Sen \cite{Sen:2008vm}. Later we describe recent developments on the computation of the $AdS_2$ path integral using supersymmetric localization. 

The quantum entropy function is a proposal based on the $AdS_2/CFT_1$ correspondence that relates the quantum degeneracy $d(q)$ of an extremal black hole with charges $q$ to a  string theory path integral on $AdS_2$, that is,
\begin{equation}\label{Z_AdS2}
d(q)=\langle e^{-iq\oint A}\rangle_{AdS_2}
\end{equation}
The path integral is performed in euclidean $AdS_2$ \footnote{We do the Wick rotation $t\rightarrow -i\theta$. In this case  the path integral configurations are weighted with $e^{S}$ with $S$ the action; it is the renormalized action that is damping the path integral, which explains the unusual sign of the exponential $e^{S}$.} and the Wilson line insertions are required  to assure that the correct boundary conditions are preserved. On $AdS_2$ we fix the electric fields and integrate instead over the chemical potentials which are the normalizable modes. We denote the Wilson line path integral (\ref{Z_AdS2}) simply by $Z_{AdS_2}$.

This formalism reduces to Wald's formalism in the limit of low curvatures or large horizon radius. That is, in a saddle point approximation we can write $Z_{AdS_2}$ as the contribution of the on-shell configuration,
\begin{equation}\label{RenEntropyFnct}
\langle e^{-iq\oint A}\rangle_{AdS_2}\simeq \text{Ren}\lbrace e^{-2\pi q e(r_0-1)+(r_0-1)2\pi \mathcal{L}(v,e,\Phi)}\rbrace=e^{2\pi q e-2\pi \mathcal{L}(v,e,\Phi)}
\end{equation}where $r_0$ is an IR cuttoff \cite{Sen:2009vz} and $\text{Ren}$ denotes renormalization by appropriate boundary counter terms. The most R.H.S expression is the exponential of Wald's entropy or equivalently Sen's entropy function \cite{Sen:2005wa}, which in this context is interpreted as the on-shell renormalized action on $AdS_2$. 

Additional quantum corrections can be systematically computed in perturbation theory. For example, in \cite{Banerjee:2010qc,Sen:2011ba,Banerjee:2011jp,Sen:2011cj} logarithmic corrections to the entropy are computed  by integrating out the massless fields \footnote{Similar logarithmic corrections are found in \cite{Keeler:2014bra,Larsen:2014bqa} using properties of the supersymmetry algebra.}.  These are found to be in agreement with the microscopic answers described in sections \S\ref{micro N=8 sec}  and \S\ref{micro N=4 sec}.

\subsection{$\mathcal{N}=4$ supergravity and attractor background}\label{N=4 sugra sec}

In this section we describe the on-shell attractor background of one-quarter BPS black holes in four dimensional $\mathcal{N}=4$ supergravity. This includes the CHL compactifications in both $K3$ and $T^4$ compactifications. Along with this, we describe how the one-eighth BPS black hole can be embedded in $\mathcal{N}=4$ supergravity. Additional details can be found in \cite{Sen:2007qy} and references therein.

As explained in section \S\ref{micro N=4 sec},  CHL compactifications are  $\mathbb{Z}_N$ orbifolds of IIB string theory on $M\times \tilde{S}^1\times S^1$, where $M$ is either $K3$ or $T^4$, that preserve $\CN=4$ supersymmetry.   After the orbifold the U-duality group is reduced to an $SL(2,\mathbb{Z})\times SO(6,r-6;\mathbb{Z})$ subgroup.

The massless bosonic spectrum consists of the  string metric $g_{\mu\nu}$, $r=2k+8$ abelian gauge fields $A^i_{\mu},\,(i=1\ldots r)$ with $k$ given by (\ref{N and k}), the axion-dilaton field $a+iS$ and a set of $r\times r$ matrix valued scalar fields $M$ subject to the constraint $MLM^T=L,\;M^T=M$ with $L$ the $SO(6,r-6)$ T-duality invariant matrix. In terms of supermultiplets we have the $\mathcal{N}=4$ supergravity multiplet, which contains the metric, six $U(1)$ gauge fields, the axion-dilaton $a+iS$ and the fermionic superpartners, interacting with $2k+2\,$ $\mathcal{N}=4$ vectormultiplets each containing a $U(1)$ gauge field, six scalars and corresponding fermionic superpartners.

At the two derivative level the four dimensional effective Lagrangian is the same for all the compactifications mentioned, with the exception of the number of $U(1)$ gauge fields. Hence, at this order in derivatives, we can study the solution that extremizes the quantum entropy functional (\ref{Z_AdS2}) in quite generality.  This configuration preserves all the symmetries of $AdS_2\times S^2$ and for this reason it has the form
\begin{eqnarray}\label{Het frame attrt background}
 &&ds^2=\frac{\vartheta_1}{8}\left(-(r^2-1)dt^2+\frac{dr^2}{r^2-1}\right)+\frac{\vartheta_2}{8}(d\theta^2+\sin^2(\theta)d\phi^2)\\
&&S=u_s,\;a=u_a,\;M=u_{ij},\\
&&F^i_{rt}=\frac{e^i}{4},\;F^i_{\theta\phi}=\frac{p^i}{16\pi}\sin(\theta)
\end{eqnarray}
with constant fields $v_1,\,v_2,\,u_s,\,u_a,\,u_{ij}$. Here $F_{\mu\nu}$ is the field strength of the abelian gauge field with $e$ the electric field and $p$ the magnetic charge, which are also constant. 

After substituting the on-shell values of the electric fields $e^i$ and renormalizing the action as in (\ref{RenEntropyFnct}), the entropy functional for two derivative supergravity has the form
\begin{equation}\label{entropy fnct}
 \text{Ren}(S)=\frac{\pi}{2}\left[2u_s(\vartheta_2-\vartheta_1)+\frac{\vartheta_1}{\vartheta_2}\frac{|Q-\tau P|^2}{u_s}\right]
\end{equation}with $\tau=u_a+iu_s$ and $Q,\,P$ are respectively the electric and magnetic charge vectors (in the absolute square on the RHS of (\ref{entropy fnct}) it is implicit a contraction with the T-duality $SO(6,r-6)$ invariant matrix $L$). We have used $P^i=L_{ij}p_j/4\pi$ and $Q_i=2q_i$, with $q_i$ the electric charge associated with the electric field $e^i$ \cite{Sen:2007qy}. With these definitions, T-duality acts linearly on the charge vectors $(Q^i,P^i)$.

Further extremization with respect to $\vartheta_1,\vartheta_2$ leads to an equation that relates the size of $AdS_2\times S^2$ in terms of the moduli $\tau$, that is,
\begin{equation}\label{size attractor eq}
 \vartheta_1=\vartheta_2=\frac{|Q-\tau P|^2}{2u_s^2}.
\end{equation}
After substituting this back in (\ref{entropy fnct}) we obtain an effective entropy function for the moduli $\tau$ 
\begin{equation}\label{ent fnct sen}
 \text{Ren}(S)|_{\vartheta_1=\vartheta_2}=\frac{\pi}{2}\frac{|Q-\tau P|^2}{u_s}
\end{equation}
which in turn leads to the attractor values
\begin{equation}\label{attractor background}
 u_s=\frac{\sqrt{Q^2P^2-(Q.P)^2}}{P^2},\;u_a=\frac{Q.P}{P^2},\;\vartheta_1=\vartheta_2=P^2.
\end{equation}

The $\CN=8$ attractor is determined similarly by considering a truncation of $\mathcal{N}=8$ to $\mathcal{N}=4$ supergravity \cite{Banerjee:2011jp}. Under this truncation all the R-R and R-NS fields are set to zero, so in this case the one-eighth BPS black hole is equivalent to a one-quarter BPS black hole in $\mathcal{N}=4$ supergravity with the same charges and therefore the same attractor background (\ref{attractor background}). 

We now study the effect of higher derivative corrections. In this case the attractor background receives corrections that depend on the $\mathbb{Z}_N$ orbifold. The effective action contains Gauss-Bonnet corrections of the form \cite{Sen:2007qy}
\begin{equation}\label{high der corrct}
 \Delta S=\int d^4x \sqrt{-\text{det}G}\phi(a,S)\{R_{\mu\nu\rho\sigma}R^{\mu\nu\rho\sigma}-4R_{\mu\nu}R^{\mu\nu}+R^2\}
\end{equation}
where the curvature tensor is computed with respect to the Einstein metric $G_{\mu\nu}=Sg_{\mu\nu}$ and the function $\phi(a,S)$ has the form
\begin{equation}\label{phi_as}
 \phi(a,S)=-\frac{1}{64\pi^2}\left[(k+2)\ln(S)+\ln g(a+iS)+\ln g(-a+iS)\right]+\text{constant}
\end{equation}with $g(\tau)$ given by
\begin{equation}
 g(\tau)=e^{2\pi i \alpha\tau}\prod^{\infty}_{n=1}\prod_{r=0}^{N-1}(1-e^{2\pi i r/N}e^{2\pi i n \tau})^{s_r}.
\end{equation}Here $s_r$ counts the number of harmonic $p$-forms of $M$ with $\tilde{g}$ eigenvalue $e^{2\pi i r/N}$ weighted with $(-1)^p$ and $\alpha=1,0$ for $M=K3,\, T^4$ respectively. The function $g(\tau)$ is a modular function of weight $k+2$ under a $\Gamma_1(N)$ congruence subgroup and it has the form already described in \S \ref{micro N=4 sec}, formulas (\ref{K3 g(t)}) and (\ref{T4 g(t)}).

In the $AdS_2$ formalism, the renormalized on-shell action is computed from a local and analytic Lagrangian. However terms as $\ln S$ in (\ref{phi_as}) are non-analytic and thus, in computing the entropy functional we throw away such terms. The attractor equations for $\vartheta_1,\vartheta_2$ (\ref{size attractor eq}) are not modified by the higher derivative corrections which leads as before to an effective entropy functional for the moduli $\tau$, that is,
\begin{equation}\label{entrpy fnct higher der}
  \text{Ren}(S)=\frac{\pi}{2}\left[\frac{|Q-\tau P|^2}{u_s}-\Omega(\tau,\bar{\tau})\right],
\end{equation}
with
\begin{equation}
 \Omega(\tau,\bar{\tau})=\ln g(\tau)+\ln g(-\bar{\tau}).
\end{equation}
The values of $u_s,u_a$ are determined by extremization and this leads to  
 \begin{equation}\label{size high der corrct}
 \vartheta_1=\vartheta_2=P^2-\partial_{u_s}\Omega(\tau,\bar{\tau})
\end{equation}
For large charges, $u_s$ is very large and equation (\ref{size high der corrct}) approximates to
\begin{equation}\label{size inst corrections}
 \vartheta_1=P^2+4\alpha+\mathcal{O}(e^{-2\pi u_s}).
\end{equation}
In the $\CN=8$ case the higher derivative terms (\ref{high der corrct}) are absent and thus the result (\ref{attractor background}) is exact up to this order in alpha prime.

In the following we use a six dimensional description of the attractor background to highlight the geometrical nature of the moduli $\tau$. 

A detailed description of the U-duality map can be found for example in \cite{Dabholkar:2005dt}. This requires mapping the Heterotic configuration to a mixed NSNS RR configuration in IIA on $K3\times S^1\times \tilde{S}^1$ and then do a series of M-theory lift/reduction and T-duality transformations to land on IIB on $K3\times S^1\times S^1_M$ where $S^1_M$ is the M-theory circle in the IIA description. This leads to a configuration with $q_1$ units of momentum along the circle $S^1_M$, $p^1$ units of KK monopole associated with the same circle and $q_0$ units of momentum along the circle $S^1$. On the other hand a $(q_a|p^a)$ configuration in IIA maps to D3-branes wrapping cycles $S^1_M\times \gamma^a$ and $S^1\times \tilde{\gamma}^a$ respectively with $\gamma^a\in H^2(K3)$ and $\tilde{\gamma}^a$ is the Poincare dual of $\gamma^a$. The 
$(q_0,q_1|p^1)$ configuration corresponds geometrically to a local $AdS_3\times S^3$ metric
\begin{equation}\label{local AdS3xS3}
 ds^2_6=\frac{\vartheta}{4}ds^2_{AdS_2\times S^2}+\frac{\vartheta}{4(e^0)^2}(dy-e^0(r-1)dt)^2+\frac{\vartheta}{4(p^1)^2}\left(dz+\frac{e^1}{e^0}dy+p^1\cos \theta d\phi\right)^2
\end{equation}
where $ds^2_{AdS_2\times S^2}$ has unit size. The circles $S^1$ and $S^1_M$ correspond respectively  to the $y$ and $z$ directions and the parameters $e^{0,1}$ are four dimensional electric fields associated with the charges $q^{0,1}$ respectively. They parametrize a torus with metric
\begin{equation}\label{torus}
 ds^2=\frac{\vartheta}{4(e^0)^2}dy^2+\frac{\vartheta}{4(p^1)^2}\left(dz+\frac{e^1}{e^0}dy\right)^2
\end{equation}
with complex structure $\tau$ and volume given by
\begin{equation}\label{tau T^2}
 \tau=e^1/e^0+ip^1/e^0,\;\text{vol}(T^2)=\frac{\vartheta}{4p^1 e^0}=\frac{P^2}{8p^1 e^0}.
\end{equation}
where we have used the attractor values (\ref{attractor background}). The geometry (\ref{local AdS3xS3}) being locally $AdS_3\times S^3$, preserves an $SL(2,\mathbb{R})_R\times U(1)_L\times SU(2)_R\times U(1)_L$ subgroup of the full $SL(2,\mathbb{R})_R\times SL(2,\mathbb{R})_L\times SU(2)_R\times SU(2)_L$ isometry of global $AdS_3\times S^3$. 

To arrive at the geometry (\ref{local AdS3xS3}) we uplifted the four dimensional IIB configuration first to five dimensions as explained for example in \cite{Beasley:2006us,deWit:2009de} and then to six dimensions by analogy. The four dimensional gauge fields associated with the charges $q^0,q^1$ are respectively
\begin{equation}
A^0_{4d}=-e^0(r-1)dt,\;A^1_{4d}=e^1(r-1)dt+p^1\cos(\theta)d\phi
\end{equation}
where $A^1$ is defined only locally due to a Dirac monopole singularity. Under this process we take $A^0$ as the Kaluza-Klein gauge field associated with the circle $S^1$ and write the five dimensional gauge field, following \cite{deWit:2009de}, as
\begin{eqnarray}
A^1_{5d}&=&\frac{e^1}{e^0}(dy-e^0(r-1)dt)+e^1(r-1)dt+p^1\cos(\theta)d\phi\nonumber\\
&=&\frac{e^1}{e^0}dy+p^1\cos \theta d\phi.
\end{eqnarray}
The uplift to six dimensions is done by analogy by turning $A^1_{5d}$ into the Kaluza-Klein gauge field associated to the circle $z$.

\subsection{Localization in $\mathcal{N}=2$ Off-shell Supergravity}\label{Loc Sugra sec}

In this section we review recent developments on the exact computation of the  $AdS_2$ path integral  using supersymmetric localization. For more details we refer the reader to \cite{Dabholkar:2011ec,Dabholkar:2010uh,Banerjee:2009af,Gomes:2013cca} and references therein.

Supersymmetric localization can be explained succinctly as follows. In supersymmetric QFT's we introduce a regulator in the action of the form $QV$ with  $Q$ an hermitian supercharge that generates a $U(1)$ symmetry and $V$ is a deformation invariant under that $U(1)$. Because both the action and the deformation are annihilated by the supercharge, the path integral does not change and hence $V$ generates an equivalence class of Lagrangians. In mathematical terms the Lagrangians are equivariantly cohomologous.   In other words, we have the following identity
\begin{equation}
\int e^{-S}=\int e^{-S-tQV}
\end{equation}
for any $t$, with $S$ the physical action. By choosing  $V$ for which the deformation $QV$ is positive semidefinite, the limit $t\rightarrow +\infty$ leads to a drastic simplification: the path integral localizes over the saddles of the deformation $QV$ and the one-loop approximation becomes exact at these points. That is,
\begin{equation}\label{localization exact deformation}
\int e^{-S}=\sum_{\sigma\in\text{ saddles }QV}e^{-S(\sigma)}\times Z^{QV}_{1\text{-loop}}
\end{equation}
where $Z^{QV}_{1\text{-loop}}$ is a superdeterminant that depends only on the choice of the deformation and not on particular details of the physical theory. 

In supergravity, localization is technically more challenging. The main difficulty comes from the fact that we have to deal with local supersymmetry and so it is not clear how to translate the equivariant localization principle to this context.  It is  possible that the $AdS_2$ path integral  receives contributions from only backgrounds that preserve a certain $U(1)$ symmetry as argued in \cite{Banerjee:2009af}, thus allowing for the equivariant principle to take place.

 Despite these difficulties, the authors in \cite{Dabholkar:2010uh} considered  supersymmetric localization in $\CN=2$ off-shell supergravity. The analysis focused only on off-shell vectormultiplets living on a rigid $AdS_2\times S^2$ geometry. In this background, we can find a supercharge $Q$ that squares to the combination $L_0-J_0$, where $L_0$ and $J_0$ are rotations on $AdS_2$  and $S^2$ respectively, and therefore we can use that supercharge to localise. Under the principle described before, the path integral localizes over a $n_V+1$-dimensional space of configurations, with $n_V$  the number of $\CN=2$ vector multiplets in the theory \footnote{In $\CN=2$ superconformal off-shell supergravity we need to introduce a compensating vectormultiplet since the Weyl multiplet does not carry any vector. }. This is possible because the localization equations allow for  the vectormultiplet scalars $X^I$ to have a non-trivial solution at the cost of turning on the auxiliary fields $Y^I$. In terms of radial coordinates\footnote{We take the $AdS_2$ metric to be $ds^2=dr^2/(r^2-1)+(r^2-1)d\theta^2$.} they have the off-shell profile  
\begin{equation}\label{localization solution}
 X^I=X^{I*}+\frac{C^I}{r},\;Y^I=\frac{2C^I}{r^2},\;I=0\ldots n_V
\end{equation}with $C^I$ an integration constant and $X^{I*}$ is the attractor value. By (\ref{localization exact deformation}) we are instructed to compute the supergravity action on these solutions, properly renormalized, and integrate over the constant $C^I$'s. We obtain 
\begin{equation}\label{loc partition fnct}
 Z_{AdS_2\times S^2}=\int \prod_{I=0}^{n_V+1}d\phi^Ie^{-\pi q_I\phi^I +\mathcal{F}(\phi,p)},
\end{equation}where $\mathcal{F}(\phi,p)$ is a function of the prepotential $F(X)$ and magnetic charges $p^I$
\begin{equation}\label{ren action prepotential}
 \mathcal{F}(\phi,p)=-2\pi i\left[F\left(\frac{\phi^I+ip^I}{2}\right)-\bar{F}\left(\frac{\phi^I-ip^I}{2}\right)\right].
\end{equation}
and $\phi$ is a certain combination of $C^I$ in (\ref{localization solution}) and the attractor values. It is also possible to show that the contribution of a large class of D-term type corrections in off-shell supergravity vanishes exactly on the localization solution \cite{Murthy:2013xpa}.
 
Furthermore, since the theory is abelian, the $QV$ deformation is purely quadratic \cite{Dabholkar:2011ec,Dabholkar:2010uh}. This implies that the one-loop determinants cannot have any dependence on the constant $C^I$'s. Instead, the only dependence comes from the parameter $\vartheta$, which is the size of $AdS_2\times S^2$.  In this case, since both the action and the deformation are scale invariant this is possible due to an anomalous scaling of the one-loop determinants (\ref{localization exact deformation}). 
 
 To understand the scaling properties of the integral (\ref{loc partition fnct}) consider the following. The integration variable $\phi^I$ in (\ref{loc partition fnct}) has the scale invariant form \footnote{Note that in \cite{Dabholkar:2010uh} it was used a gauge with $\omega=1$. } 
\begin{equation}\label{phi scale invariant}
 \phi^I=e^I+2\omega^{-1}C^I=\text{Re}(2\omega^{-1}X^I)
\end{equation}where $\omega$ is defined in relation to the size of $AdS_2$ as
\begin{equation}
 \vartheta=\frac{1}{\omega^2}.
\end{equation}
In the $\CN=2$ off-shell formalism $\omega$ is proportional to the attractor value of the auxiliary tensor $T_{\mu\nu}$ which in the on-shell theory gives rise to the graviphoton field.
Furthermore, the renormalized action depends on $C^I$ only via the scale invariant combination (\ref{phi scale invariant}), while the charges $q,p$ are scale invariant by definition. Therefore, from the anomalous scale transformation of the partition function
\begin{equation}\label{scale anomaly}
Z(\lambda\omega,\lambda X)=\lambda^{-2\beta}Z(\omega,X)
\end{equation}
we conclude that the one-loop determinants must give a factor of $\vartheta^{\beta}$, with $\vartheta$ the size of $AdS_2\times S^2$ and $\beta$  is the scale anomaly of the vector-multiplet partition function. Later we compute $\beta$ for the case of interest.

At this point it is important to make a few remarks concerning the one-loop computation of  \cite{Murthy:2015yfa,Gupta:2015gga}. Their approach is different since they first consider a computation where the metric is rigid with a charge independent constant size and then claim the result to be valid also when the metric is fluctuating, by means of a Weyl transformation, which is seen as a gauge choice  \footnote{In a gauge where the Kahler potential $e^{-\mathcal{K}}(X)=i(F_I\bar{X}^I-\bar{F_I}X^I)=1$ the supergravity Lagrangian has the canonical Einstein-Hilbert term and therefore the conformal factor of the metric is effectively fluctuating.}. In order to do the computation with a fluctuating metric it is strictly necessary to consider the Weyl multiplet coupled to the remaining matter multiplets. However, it is an open problem how to define an equivariant complex using local supersymmetry. A few comments on this problem can be found in \cite{Gomes:2013cca}.

\section{Supersymmetry and Measure}\label{measure sec}

In this section we derive the exact measure for the finite dimensional integral  (\ref{loc partition fnct}). 

There are two main tasks underlying the derivation. First we argue that the measure for the $\CN=4$ vector-multiplets, that is, the measure for the variables $\phi^a$ with $a=2\ldots N_V+1$, is flat. To justify this we  consider first the theory of free $N_V$ $\CN=4$ vector-multiplets living on a rigid $AdS_2\times S^2$ and compute the exact partition function. This will allow us to fix the measure. Finally we consider the contribution of the $\CN=4$ supergravity multiplet. Using a supersymmetric Chern-Simons theory on $AdS_2\times S^1$ we determine a one-loop dependence on the background metric. This will fix the measure for the variables $\phi^0$ and $\phi^1$ in  (\ref{loc partition fnct}) that parametrize normalizable fluctuations of the axion-dilaton in the $\CN=4$ supergravity multiplet.  

Schematically the $\CN=4$ answer has the form 
\begin{equation}\label{measure sec loc int}
 d(q,p)_{\CN=4}=\frac{1}{C}\int \prod_{a=0}^{N_V+1}d\phi^a \mathcal{M}_{1/4}(\phi,p) \exp{\left[-\pi q_I\phi^I +\mathcal{F}^{(0)}(\phi,p)\right]}+\sum_{\text{instanton}}\int \prod_{a=0}^{N_V+1}d\phi^a\ldots
\end{equation}
where $\mathcal{F}^{(0)}(\phi,p)$ is defined as (\ref{ren action prepotential}) with the zero instanton $\mathcal{N}=2$ prepotential
\begin{equation}\label{zero prepotential}
 F(X)=\frac{X^1}{X^0}C_{ab}X^aX^b+\omega^2\, c_1\,\frac{X^1}{X^0}
\end{equation}
that describes the coupling of $N_V$ $\mathcal{N}=4$ vector multiplets to the supergravity multiplet. Here $c_1=0,1$ for $T^4$  and $K^3$ models respectively. The factor $\mathcal{M}_{1/4}(\phi,p)$ is the effective measure for the supergravity multiplet fields that we want to determine. 

% The measure has the exact form
% \begin{equation}\label{sugramultiplet measure}
%  \mathcal{M}_{1/4}=\frac{P^2+4c_1}{\phi^0p^1}
% \end{equation}

Furthermore, we argue that there are subleading saddle points, the second term in (\ref{measure sec loc int}), which can be interpreted as instanton contributions to the $AdS_2$ path integral. Based on the localization integral (\ref{loc partition fnct}) with the non-perturbative prepotential we suggest a reinterpretation of the instanton contributions in terms of a renormalization of the Chern-Simons couplings. Proceeding analogously to the zero instanton case we obtain an unitarity condition that truncates the instanton sum leading precisely to the tail of Bessel functions found in (\ref{finite sum Bessels}).

For $\CN=8$ black holes we use a  $\CN=4$ truncation. For this reason the exact answer still has the form of (\ref{measure sec loc int}), but in this case the instanton contributions are not present, with the prepotential having the tree level form $F(X)=X^1/X^0C_{ab}X^aX^b$. However, from this point of view there is an additional contribution coming from the fields that are thrown away under the truncation of $\CN=8$ supergravity. We denote this contribution by $Z_{\text{odd}}$. The final answer has the form
\begin{equation}
 d(q,p)_{\CN=8}=\frac{1}{C}\int \prod_{a=0}^{N_V+1}d\phi^a \,\mathcal{M}_{1/8}(\phi,p)\,\exp{\left[-\pi q_I\phi^I +\mathcal{F}(\phi,p)\right]}\times Z_{\text{odd}}
\end{equation}
with the measure $\mathcal{M}_{1/8}(\phi,p)$ and $Z_{\text{odd}}$ the quantities we want to determine. Here $N_V=6$ is the number of $\CN=4$ vector-multiplets of the truncation and $C$ is a normalization constant.

\subsection{Odd fields contribution}

In \S \ref{N=4 sugra sec} we explained a truncation of $\CN=8$ to $\CN=4$ supergravity. This truncation consists in setting all the RR and RNS fields to zero in IIB string theory. From the string worldsheet this is equivalent to consider only the fields which are even under $(-1)^{F_L}$, with $F_L$ the left fermion number, in a sector invariant under a right $\CN=4$ subalgebra. This is the sector where the microscopic formula (\ref{deg bessel 1/8}) is valid. The $\CN=4$ truncation consists of the supergravity multiplet together with six vector-multiplets and the U-duality group is reduced to a $SL(2,\mathbb{Z})\times SO(6,6;\mathbb{Z})$ subgroup. 

The computation is based on the assumption  that the contribution from  even and odd fields factorize. This is justified in part from the fact that at the quadratic level the fluctuations over the odd fields \cite{Banerjee:2011jp} does not mix with the fluctuations of the even fields due to the symmetry under $(-1)^{F_L}$ and therefore they can be integrated out to obtain an effective $\CN=4$ supergravity. It is possible that for the class of BPS states we are interested in, namely those which are invariant under the right $\CN=4$ subalgebra, the factorization is exact.

 On the other hand, the microscopic counting formulas for the CHL models on $T^4$ compactifications strongly suggest that this is the case. Namely the $T^4\times T^2/\mathbb{Z}_2$ orbifold has precisely the same massless spectrum and the same U-duality group $SL(2,\mathbb{Z})\times SO(6,6;\mathbb{Z})$ as the $\CN=4$ truncation. For large charges the leading microscopic degeneracy has the form
 \begin{equation}
 d(Q,P)|_{T^4/\mathbb{Z}_2}\simeq 2(-1)^{Q.P+1}\int_{\hat{\mathcal{C}}} \frac{d^2\tau}{\tau_2^{5}}\pi P^2 \exp{\left[\frac{\pi}{2}\frac{|Q+\tau P|^2}{\tau_2}\right]}
\end{equation}
which is precisely the $\CN=8$ answer  (\ref{deg 1/8 integral}) up to a factor of $(P^2)^{-4}$. This suggests that the odd fields contribution should be given by
\begin{equation}\label{Zodd}
Z_{\text{odd}}=\frac{1}{(P^2)^{4}}
\end{equation}

  At the quadratic level it is easy to compute (\ref{Zodd}). The gaussian integrals will give factors of the dilaton and $\vartheta$, the size of $AdS_2$, since these are the only parameters available. The dependence on the first can be determined by noting that at the quadratic level the supergravity action has a symmetry \cite{Sen:1995in,Sen:2009bm} in which the dilaton is shifted by $\ln\lambda^{-1}$, the NS fields remain invariant and the RR fields are multiplied by $\lambda$. This implies that the quadratic action for the odd fields does not have dependence on the dilaton and thus the zero mode argument of section \S\ref{vec measure sec} gives a trivial answer. On the other hand, there is a non-trivial dependence on $\vartheta$, which equals $P^2$ in this case, due to a scaling anomaly. This was computed in \cite{Banerjee:2011jp} using the heat kernel method. It is found a contribution of $-4\ln P^2$ to the effective action coming from the odd fields. This leads to the result (\ref{Zodd}) as we wanted.

\subsection{Vectormultiplet measure}\label{vec measure sec}

To understand the measure for the vectormultiplets we consider first the theory of $N_V$ free $U(1)$ $\mathcal{N}=4$ vector multiplets on $AdS_2\times S^2$ with microcanonical boundary conditions, that is, we fix the electric fields and allow for the chemical potentials to fluctuate. To have a variational problem consistent with these boundary conditions  we need to insert appropriate Wilson lines. In addition, we need to add boundary counter terms to remove IR divergences. 

The $\CN=4$ Lagrangian contains a Maxwell and a theta term
\begin{equation}\label{free maxwell}
 \frac{1}{g^2}\int F\wedge \star F+\theta \int F\wedge F,
\end{equation}
 with $g$ the $U(1)$ coupling constant,  together with six scalars and respective fermionic terms. Without loss of generality we set the theta term to zero for the moment. Since the $U(1)$ theory is free the partition function is semiclassically exact and in this case we only need to know the on-shell action and the one-loop determinants of fluctuations. By comparing this with the localization computation we will find evidence for the measure used. Note that in the localization computation the one-loop determinants $Z^{QV}_{1\text{-loop}}$ (\ref{localization exact deformation}) do not depend on particular details of the physical theory and therefore they can be computed in the case of the free theory.

 We follow closely Witten's work on free Maxwell theory on a compact manifold \cite{Witten:1995gf} and highlight the main differences for the non-compact case. As explained, in this case the semiclassical approximation is exact and thus the partition function is a sum over saddles times a determinant over fluctuations.  Before going in details about the computation, we present the final expression for the partition function: 
\begin{equation}\label{Z N=4 free vector}
 Z_{\mathcal{N}=4}^{\text{free}}=e^{\frac{\pi}{2}(q,q)g^2+\frac{\pi}{2}\frac{1}{g^2}(p,p)}g^{N_V}
\end{equation}
where $(\,,\,)$ is a measure induced by the hodge  $\star$-operator via the kinetic terms (\ref{free maxwell}) and $q$ and $p$ are respectively the electric and magnetic charge vectors. The exponential term  comes from the evaluation of the renormalized on-shell action on solutions that carry electric and magnetic charges, that is, which have field strength $F=e\,\text{vol}_{AdS_2}+p\,\text{vol}_{S^2}$. On the other hand, the factor $g^{N_V}$ comes from the one-loop determinant of fluctuations whose computation is more involved. Roughly, there is a factor of $g$ for each non-zero mode fluctuation times a zeta-function regularized super-determinant that depends only on the size of $AdS_2$. The super-determinant can be computed as in \cite{Banerjee:2010qc} using the heat kernel method- it is found that each $\CN=4$ vector-multiplet gives a trivial contribution- while the dependence on $g$ is determined using an ultralocality argument. Following \cite{Witten:1995gf}, we introduce a regulator to count the number of non-zero modes. This number is then equal to the total number of modes, which is a  local quantity, minus the number of zero modes; the first can be renormalized to one by adding appropriate local counter terms and therefore we are left with $g^{-N_{\text{zero}}}$ where $N_{\text{zero}}$ is the number of zero modes. The main difference with the computation \cite{Witten:1995gf} is that here we obtain an infinite number of zero modes since the space is non-compact.

Lets analyse the zero mode contribution. In $AdS_2$ only the $U(1)$ gauge fields can have zero modes \cite{Camporesi:1994ga,Banerjee:2010qc}. These correspond to gauge transformations with non-normalizable parameters, that is,
\begin{equation}\label{U1 zero mode}
A^{\text{zero}}_l=d\Phi_l
\end{equation}
with 
\begin{equation}\label{U1 zero mode 2}
\Phi_l=\frac{1}{\sqrt{2\pi|l|}}\left[\frac{\sinh(\eta)}{1+\cosh(\eta)}\right]^{|l|}e^{i\theta l},\;l=\pm1,\pm2,\ldots
\end{equation}
where we have used euclidean hyperbolic metric for $AdS_2$ \footnote{In these coordinates the metric is $ds^2=d\eta^2+\sinh(\eta)^2d\theta^2$}.
The mode $\Phi_l$ is not normalizable on $AdS_2$ since as we approach the boundary $\eta\sim \infty$, it behaves as $\Phi_l\sim e^{i\theta l}$ and thus its norm diverges. On the other hand, the gauge field (\ref{U1 zero mode}) respects the microcanonical boundary conditions and has a finite squared norm. Note that this field configuration cannot be gauged away because $\Phi_l$ is not normalizable, and so $A^{\text{zero}}_l=d\Phi_l$ is indeed a zero mode.

To compute the regularized number of non-zero modes we use the normalization condition \cite{Banerjee:2010qc}
\begin{equation}\label{ultralocal norm}
\int D[A_{\mu}(x)]e^{-\frac{1}{g^2}\int A\wedge \star A}=1.
\end{equation}
Basically this means that any local  quantity of $A_{\mu}(x)$ can be renormalized to one by adding local counter terms.
Procceding as in \cite{Witten:1995gf} the one-loop determinant gives $g$ to the power of the number of non-zero modes. Multiplying and diving by $g^{N_{\text{zero}}}$, with $N_{\text{zero}}$ the number of zero modes, the normalization (\ref{ultralocal norm}) leads to the result 
 \begin{equation}\label{non-compact 1-loop}
  1\text{-loop}\sim g^{-N_{\text{zero}}}.
 \end{equation}

 In general the number of zero modes in compact manifolds is finite, however, because we are on  $AdS_2$ this  number is infinite- see equations  (\ref{U1 zero mode}) and (\ref{U1 zero mode 2}). In this case the infinity is due to an IR  rather than an UV divergence and hence it can be renormalized by introducing boundary counter terms. Proceeding as in \cite{Banerjee:2010qc} we obtain
\begin{eqnarray}\label{renorm zero mode dim}
N^{\text{zero}}&=&\sum_{l}1=\sum_l(d\Phi_l,d\Phi_l)=\cosh(\eta_0)-1\\
&\substack{=\\\text{reg}}&-1
\end{eqnarray}
where we have removed the cuttoff dependent term $\cosh(\eta_0)$.
The final answer for the one-loop determinant is therefore
\begin{equation}
Z^{\text{free}}_{1\text{-loop}}=g^{-N^{\text{zero}}_{\text{reg}}\times N_V}=g^{N_V},
\end{equation}
as we wanted to show.

At this point it is useful to make a comparison with the compact case. For simplicity lets consider a compact manifold with constant positive curvature like the four-sphere. In this case there are no fermionic zero modes because the square of the Dirac operator  is positive definite by $\slashed\nabla^2\psi=R\psi$ \cite{Pestun:2007rz}, with $R$ the curvature. Furthermore, the scalars couple conformally to the curvature scalar and this generates an effective potential which sets them to zero. Therefore, the only zero modes present come from the $U(1)$ gauge fields fluctuations.  

The counting of zero modes goes as follows. In the compact case there are $b_1$ zero modes to the vector laplacian and a gauge fixing ghost zero mode due to  the constant gauge transformation that acts trivially. Hence the number of zero modes in the gauge fixed theory is precisely $b_1-b_0$, where  the minus sign is due to the ghost zero mode which is fermionic; the numbers $b_0,\,b_1$ can be identified respectively with the dimensions of the De Rham cohomology groups $H^0,\,H^1$. The one-loop contribution is then \cite{Witten:1995gf} 
\begin{equation}\label{Z-1loop compact}
Z_{1\text{-loop}}=g^{b_0-b_1}.
\end{equation}
Suppose we have $b_1=0$ as in the $AdS_2\times S^2$ case, then, since $b_0=1$, we find a factor of $g$ for the one-loop contribution (\ref{Z-1loop compact}) which is precisely the result found in the non-compact case for one vector-multiplet. 

It is an instructive exercise to write the non-compact result (\ref{non-compact 1-loop}) in a similar language, that is, in terms of cohomology. To do that we denote by $\hat{H}^1$ the space of normalizable closed  one-forms modulo exact one-forms $d\alpha$ with normalizable $\alpha$ on $AdS_2\times S^2$. Analogously we define $\hat{H}^0$ to be the space of normalizable closed  zero-forms. The dimensions of these spaces, respectively $\hat{b}^{1,0}$, must be renormalized as in (\ref{renorm zero mode dim}). Thus, on $AdS_2\times S^2$ we find $\hat{b}^0=0$ because any constant is non-normalizable, and $\hat{b}^1=-1$ by (\ref{renorm zero mode dim}). This leads to the same result (\ref{Z-1loop compact}).

Given the semiclassical computation, we turn gears to the computation of (\ref{Z N=4 free vector}) using localization. The idea is to write first the $\mathcal{N}=4$ theory in terms of $\mathcal{N}=2$ multiplets and then use the result (\ref{loc partition fnct}) for localization of $\CN=2$ supergravity. We want to show that the localization computation gives the same result as in the semiclassical approach.

The $\mathcal{N}=4$ vector-multiplet decomposes into a $\mathcal{N}=2$ vector-multiplet together with an hyper-multiplet. A theory of $N_V$ free vector-multiplets can be described by a $\mathcal{N}=2$ prepotential of the form 
\begin{equation}
 F_{free}(X)=-\frac{i}{g^2}\sum_{a,b=1}^{n_v}m_{ab}X^aX^b
\end{equation}
with $m_{ab}$ some constant matrix and $g$ is the coupling constant. 

The measure for the $\CN=4$ vectormultiplet fields in the path integral is flat because of supersymmetry. The same should be true for the full supergravity path integral. It is difficult, a priori, to find a non-flat pointwise measure that is supersymmetric and so the flat choice is the most natural. There can be the confusion, however, that the kinetic terms in the physical Lagrangian determine the measure. This is not so in the presence of supersymmetry. A good example is supersymmetric quantum mechanics. In this case we have an interacting one dimensional supersymmetric sigma model that describes a superparticle moving in a non-trivial manifold; the kinetic terms are non-tivial functions of the scalar fields (the position of the superparticle) and are given in terms of the components of the background manifold. A flat measure for the partition function leads to the correct results for the expected index theorems \cite{Niemi:1993ia,Friedan:1983xr}. Similarly there are the examples of topological sigma models \cite{AlvarezGaume:1985xfa,AlvarezGaume:1985ww} just to mention a few.

Using the result (\ref{loc partition fnct}) we find that  the partition function for $N_V$ vector-multiplets is given by the gaussian integral
\begin{equation}\label{Z free localization}
 Z_{\mathcal{N}=4}^{\text{vec}}=\int \prod_{a=1}^{N_V} d\phi^ae^{-\pi q_a\phi^a-\frac{\pi}{2g^2}\sum_{a,b=1}^{N_V}m_{ab}\phi^a\phi^b+\frac{\pi}{2}\frac{(p,p)}{g^2}}Z^{QV}_{1-\text{loop}}
\end{equation}
 The factor $Z^{QV}_{1-\text{loop}}$ can be computed using an equivariant-index theorem as described in \cite{Pestun:2007rz}, though we provide a much simpler approach. As previously explained, for a $U(1)$ gauge theory the exact deformation $QV$ is purely quadratic, which implies that the one-loop determinant cannot have any dependence on $\phi^a$. Instead it is a single function of the size of $AdS_2\times S^2$ \footnote{We have used a scale invariant measure \cite{Hawking:1976ja,Gibbons:1978ji}.}. As explained in section \S\ref{localization sec}, we must have by equation (\ref{scale anomaly})
\begin{equation}
Z^{QV}_{1\text{-loop}}=\vartheta^{\beta}.
\end{equation} 
where $\beta$ is the scale anomaly. We can compute it in the on-shell theory as in \cite{Sen:2011ba,Banerjee:2011jp}. This gives $\vartheta^{-1/12}$ for the $\CN=2$ vector and $\vartheta^{1/12}$ for the hyper, and so for the full $\CN=4$ vector-multiplet the dependence on $\vartheta$ precisely cancels. This result further agrees with the index computation of \cite{Murthy:2015yfa} if we fix the size of $AdS_2$ to be the constant $\vartheta$. Hence we find
\begin{equation}
Z_{\CN=4\,\text{vec}}^{QV}=1
\end{equation}   
Finally, integrating the gaussians in (\ref{Z free localization}) we obtain the semiclassical answer (\ref{Z N=4 free vector}) as we wanted. It is important to note that we have not introduced the induced measure used in \cite{Dabholkar:2011ec, Murthy:2015yfa,Gupta:2015gga}
\begin{equation}
 \mathcal{M}_{\text{ind}}\propto \sqrt{\text{det Im}(\partial_a\partial_b F^{free})}.
\end{equation}
Such factor would lead to additional powers of $1/g$, in disagreement with the semiclassical answer. This has implications for the one-loop determinants of the $\CN=2$ gravity and gravitino multiplets, proposed in \cite{Murthy:2015yfa}  after comparing with the on-shell results of \cite{Sen:2011ba-1}.
 
A comparison between the semiclassical result (\ref{Z N=4 free vector}) and the localization integral (\ref{Z free localization}) suggests that the modes $\phi^a$ in the integral (\ref{Z free localization}) correspond in a certain way to the zero modes of the $U(1)$ gauge fields since they both generate the one-loop factor $g^{N_V}$. This is true in the case of localization of $\CN=4$ SYM on the four-sphere \cite{Pestun:2007rz}, in which case the constant mode of the scalar that is left unfixed by the localization equations, corresponds precisely to the constant mode of the gauge transformations that gives rise to a ghost zero mode. On $AdS_2\times S^2$ we have only a partial understanding of this phenomena. It is plausible that in the presence of gravity the localization equations lift all the zero modes as originally proposed in \cite{Banerjee:2009af} and only the localization mode (\ref{localization solution}) is allowed. As a matter of fact, since the theory has an asymptotic supercharge $Q$ which squares to $L_0-J_0$ \cite{Banerjee:2009af}, with $L_0$ and $J_0$ rotations on $AdS_2$ and $S^2$ respectively, we expect the zero modes to be lifted as they have non-zero eigenvalues $L_0-J_0$. It would be interesting to understand this from the localization computation.

In the black hole problem we need to couple the $\mathcal{N}=4$ vectormultiplets to supergravity. We do this by considering a theory with prepotential 
\begin{equation}
 F(X)=-\frac{1}{2}\frac{X^1}{X^0}C_{ab}X^aX^b,
\end{equation}
with $X^1/X^0$ the axion-dilaton. The coupling constant $1/g$ is now the imaginary part of the scalar $X^1/X^0$ measured at infinity and the real part is the theta parameter. Because the prepotential is still quadratic in $X^a$, the vector-multiplet integrals  are still gaussian and thus we expect to obtain a similar  contribution  as in the free case (\ref{Z N=4 free vector}). 

Lets consider the localization solution (\ref{localization solution}) but with fixed $X^1$ and $X^0$. The supergravity path integral for this configuration subspace is
\begin{eqnarray}\label{exact pt fnct and CS}
&&\int \prod^{N_V}_{a=1}d\phi^a\exp{\left[-\pi q_0 e^0-\pi q_1 e^1-\pi q_a\phi^a+\frac{\pi}{2} \frac{p^1}{e^0}P^2-\frac{\pi}{2}\frac{p^1}{e^0}\sum_{a,b=1}^{N_V}C_{ab}\phi^a\phi^b-\pi \frac{e^1}{e^0}\sum_{a,b=1}^{N_V}C_{ab}\phi^ap^b\right]}\nonumber\\
{} \label{sugra localization integral just vecs} \\ 
 &&=\int \prod^{N_V}_{a=1}d\phi^a \exp{\left[\frac{\pi}{2} Q^2\frac{e^0}{p^1}+\pi Q.P \frac{e^1}{p^1}+\frac{\pi}{2} \frac{p^1}{e^0}P^2+\frac{\pi}{2}\frac{P^2}{p^1e^0}(e^1)^2\right]}\times\nonumber\\
&&\qquad\qquad\times \exp{\left[-\frac{\pi}{2}\frac{p^1}{e^0}\sum_{a,b=1}^{N_V}C_{ab}\left(\phi^a+e^1\frac{p^a}{p^1}+q^a\frac{e^0}{p^1}\right)\left(\phi^b+e^1\frac{p^b}{p^1}+q^b\frac{e^0}{p^1}\right)\right]}.
{}\label{sugra localization integral just vecs 2}
\end{eqnarray}
Here the electric fields $e^{0,1}$ are the on-shell values of $\text{Re}(X^{0,1})$ and $Q^2$, $P^2$ and $Q.P$ are the T-duality invariant combinations
\begin{equation}
 Q^2=-2q_0p^1+C_{ab}q^aq^b,\;\; Q.P=-q_1p^1+C_{ab}q^ap^a,\;\; P^2=C_{ab}p^ap^b,
\end{equation}
with $q_a=C_{ab}q^b$.
The last term in the exponential (\ref{sugra localization integral just vecs}) comes from a theta term  $\sim\int \theta\, F\wedge F$ with $\theta=e^1/e^0$. Take $e^1=0$ for the moment. If we perform the gaussian integrals in (\ref{sugra localization integral just vecs 2}) we obtain precisely the free answer (\ref{Z N=4 free vector}) with $1/g^2=p^1/\phi^0$. If we turn on the theta term, then the contribution from $e^1$ is still gaussian and this too can be interpreted as coming from integration over zero modes.

\subsection{Super Chern-Simons theory and gravity-multiplet measure }\label{sec CS}

By the $\text{AdS}_3/\text{CFT}_2$ holographic correspondence  \cite{Maldacena:1997re,Witten:1998qj}, we have an equality between the partition functions of string theory on $\text{AdS}_3$ and the dual $\text{CFT}_2$, that is,
\begin{equation}
Z_{\text{AdS}_3}=Z_{\text{CFT}_2}.
\end{equation}
Since $Z_{\text{CFT}_2}$ has modular properties, so does $Z_{\text{AdS}_3}$, that is, we have
\begin{equation}\label{Z_CFT2}
Z_{\text{AdS}_3}\left(\frac{a\tau+b}{c\tau+d},\frac{a\bar{\tau}+b}{c{\tau}+d}\right)=(c\tau+d)^{\omega}(c\bar{\tau}+d)^{\bar{\omega}}Z_{\text{AdS}_3}(\tau,\bar{\tau}),\;\left(\begin{array}{cc}a & b\\
																															c & d	\end{array}\right)
																															\in SL(2,\mathbb{Z})
\end{equation}
where the $\tau$ is the complex structure of the $AdS_3$ boundary torus and $SL(2,\mathbb{Z})$ is the group that parametrizes  global diffeomorphisms of this torus. Equation (\ref{Z_CFT2}) shows that the partition function is not invariant under the $SL(2,\mathbb{Z})$ action but transforms covariantly with certain weights $\omega$ and $\bar{\omega}$.  In quantum field theory language this signs an anomaly. In this section we compute this anomaly and show that it fixes the measure of the localization integral (\ref{loc partition fnct}).

To illustrate the main idea, first we study the asymptotic behaviour of the Fourier coefficients of $Z_{\text{CFT}_2}$, which we take to be holomorphic. We will find that this behaviour is fixed by two properties: the anomalous transformation of the partition function (\ref{Z_CFT2}) and the values of the polar coefficients.  

In the case of one-eighth and one-quarter BPS states the spectrum is captured by a $(0,4)$ SCFT \cite{Dabholkar:2010rm}. The partition function $Z_{\text{CFT}_2}$ is thus the elliptic genus and so it is a holomorphic object. In this case the modular relation (\ref{Z_CFT2}) is modified by the introduction of the R-symmetry chemical potential $z$. To put it another way
\begin{equation}\label{Z modular transf}
Z\left(-\frac{1}{\tau},\frac{z}{\tau}\right)=\tau^{\omega}e^{2\pi i k\frac{z^2}{\tau}}Z(\tau,z),
\end{equation}
where we only consider the transformation under the element $S\in SL(2,\mathbb{Z})$. Here $k$ is the index of the elliptic genus, which is a Jacobi form. For simplicity we set it to one.

The degeneracy of BPS states can be computed by doing an inverse Fourier transform, that is,
\begin{equation}\label{inverse fourier CFT2}
d(n,l)=\int_{0}^{1}\int_{0}^{1}d\tau dz  \, Z_{\text{CFT}_2}(\tau,z)e^{-2\pi i \tau n-2\pi i zl}.
\end{equation}
For large charges $n,l$  we can compute (\ref{inverse fourier CFT2}) by a saddle point approximation. Since the saddle is at $|\tau|\ll1$ we can use the modular property (\ref{Z modular transf}) to estimate the integral (\ref{inverse fourier CFT2}). That is, near the saddle the function $Z(-1/\tau, z/ \tau)$ is dominated by the ground state which has energy $-c/24$, with $c$ the central charge and thus by (\ref{Z modular transf}) we have
\begin{equation}
d(n,l)\simeq d_{\text{polar}}\int_{0+i/\epsilon}^{1+i/\epsilon}d\tau\int_{0}^{1} dz \,\tau^{-\omega}\exp{\left[\frac{i\pi c}{12\tau}-2\pi i\frac{z^2}{\tau}-2\pi i\tau n-2\pi i z l\right]}.
\end{equation}
where we have used a $\epsilon \gg 1$ prescription to avoid the singularity. Here the coefficient $d_{\text{polar}}$ denotes the degeneracy of the polar term. Further, we perform the $z$ integral by saddle point approximation and change variables $-i\tau=\pi c/12 t$ to obtain \footnote{We are assuming that $\omega<1/2$ in which case the integral is convergent.}
\begin{equation}\label{Bessel by AdS3}
d(n,l)\sim d_{\text{polar}}\,c^{-\omega+3/2}\,\int_{\epsilon-i\infty}^{\epsilon+i\infty}\frac{dt}{t^{-\omega+5/2 }} \exp{\left[t+\frac{\pi^2c\Delta}{24 t}\right]}
\end{equation}
with $\Delta=4n-l^2$. The range of integration only makes sense asymptotically, that is, we have extended and deformed the finite contour $t\in[0+i/\epsilon,1+i/\epsilon]$ to $\epsilon +i\mathbb{R}$. Equation (\ref{Bessel by AdS3}) shows that the number of BPS states grows with Bessel type behaviour with the index of the Bessel function determined by the weight of the modular transformation (\ref{Z modular transf}). Moreover, this Bessel comes multiplied by a factor $d_{\text{polar}}\,c^{-\omega+3/2}$ which can become important for large charges.

If in addition  $Z(\tau,z)$  has weight $\omega<1/2$, then the approximation (\ref{Bessel by AdS3}) can be completed exactly by the circle method, leading to the Rademacher expansion \cite{Dijkgraaf:2000fq}. For example, for one-eighth BPS states we have $\omega=-2$ and $c=6$. Using formula (\ref{Bessel by AdS3}) we find $d(n,l)\sim I_{7/2}(\pi \sqrt{\Delta})$ which is the exact leading result (\ref{deg bessel 1/8}) in the Rademacher expansion  \cite{Dabholkar:2011ec}.  

This exercise shows essentially two things. First, the leading asymptotics of the Fourier coefficients of $Z_{\text{CFT}_2}$ are determined by a Bessel of type $I_{\nu}(\pi \sqrt{c\Delta/6})$, with the index $\nu$ determined by the weight $\omega$ in (\ref{Z_CFT2}). This result depends only on general properties of conformal field theory and not on particular details of the theory. Moreover, it illustrates that the Bessel behaviour is intimately related to the modular anomaly (\ref{Z_CFT2}) and one can be determined from the other as shown. Finally, the asymptotic behavior contains in addition a factor of $d_{\text{polar}}\,c^{-\omega+3/2}$. Since we must have $\omega<1/2$ for convergence, this term becomes important when we are scaling the central charge to parametrically large values. This is the regime studied in the logarithmic computations of \cite{Banerjee:2010qc,Banerjee:2011jp} that we want to partially revisit here. 

In the following we determine the bulk origin of these two aspects. To do that we use a supersymmetric Chern-Simons theory on a local $\text{AdS}_3$ geometry and compute a one-loop correction to the leading saddle. We show that this correction carries an anomalous dependence on the background metric. Note that Chern-Simons theory is defined without a metric. However, quantum mechanically the dependence on the metric may be anomalous.  We use this idea to argue that the particular dependence on the background metric must be valid for not only for large but also finite charges. This determines entirely the measure for the localization integral and gives the Bessel behaviour as expected.

To compute this anomaly we consider supersymmetric Chern-Simons theory on $AdS_2\times S^1$. The Lagrangian can be obtained as a consistent truncation of six dimensional supergravity on a three sphere \cite{deBoer:1998ip,Deger:1998nm} and contains both gravitational, and abelian and non-abelian Chern-Simons terms. We use microcanonical boundary conditions that are consistent with $\text{AdS}_2$. 

This point of view is justified on general grounds of two dimensional superconformal field theories and $\text{AdS}_3$ holography. It is well known that Chern-Simons theory captures many aspects of $(0,4)$ SCFT's  not only at the on-shell level \cite{Kraus:2006nb,Kraus:2005vz,Hansen:2006wu} but also at the quantum level \cite{Dabholkar:2014ema,Dabholkar:2010rm,Dijkgraaf:2000fq}. On the other hand, the anomaly can be computed at long wavelengths depending only, as we show, on global properties of the space given by a certain cohomology structure of the gauge transformations.  

The metric of $AdS_2\times S^1$ is
\begin{equation}\label{AdS2xS1 metric}
ds^2=\frac{\vartheta}{4}\left(\sinh(\eta)^2d\theta^2+d\eta^2\right)+\frac{\vartheta}{4(\phi^{0})^2}\left(dy-i\phi^{0}(\cosh(\eta)-1)d\theta\right)^2.
\end{equation}
with $\theta,\,y$ periodically identified and $\vartheta$ is determined by the attractor background (\ref{attractor background}). Under a reduction of five dimensional supergravity on the circle $S^1$ the moduli $\phi^0$ becomes the real part of $X^0$ in the four dimensional theory \cite{Banerjee:2011ts}.  

The key aspect of the metric (\ref{AdS2xS1 metric}) that we want to explore, is the fact that it corresponds to a quotient of global $\text{AdS}_3$ \cite{Murthy:2009dq} by an additive group $\Gamma$. To see this, take first global $\text{AdS}_3$ with metric\footnote{This is actually the universal cover.}
\begin{equation}
ds^2=\cosh(\rho)dt^2+\sinh(\rho)^2d\psi^2+d\rho^2
\end{equation}
with $-\infty<t<+\infty$ and $\psi\equiv \psi+2\pi$ and then consider the identification by the group
\begin{equation}\label{ident global AdS3}
\Gamma:\;(t,\psi) \sim ( t +\frac{\pi}{\phi^0},\psi+ i\frac{\pi}{\phi^0})\sim (t, \psi+2\pi) .
\end{equation}
Under the coordinate change $t=y/(2\phi^0)$, $\eta=2\rho$ and $\psi=\theta+iy/(2\phi^0)$, the metric of global $\text{AdS}_3$ becomes precisely that of $AdS_2\times S^1$ (\ref{AdS2xS1 metric}) with $\vartheta=1$. 

More generally, we can consider the identification of points $z=\psi+it$ on the boundary $\mathbb{C}$ of the universal cover of $\text{AdS}_3$ by the additive group $z\sim z +2\pi n+2\pi m\tau$. This construction leads to a solid torus with a boundary that has complex structure $\tau$. Other identifications consist of $z\sim z +2\pi n+2\pi m(a\tau+b)/(c\tau+d)$ with integers $a,b,c,d$ obeying $ad-bc=1$. This leads to the same torus but now with complex structure $a\tau+b/c\tau+d$. The interior, on the other hand, is obtained by filing the solid torus with a diffeomorphism that changes the cycles that become contractible and non-contractible. These are the well known $SL(2,\mathbb{Z})$ family of $\text{AdS}_3$ solutions \cite{Dijkgraaf:2000fq,Maldacena:1998bw}. The space with $AdS_2\times S^1$ metric (\ref{AdS2xS1 metric})  corresponds to a quotient with Lorentzian $\tau=i \phi^0$ and $\bar{\tau}\rightarrow \infty$ \footnote{In the Lorentzian version $\tau$ and $\bar{\tau}$ are independent variables.} with an $SL(2,\mathbb{Z})$ filling corresponding to $a=d=0$ and $c=-b=1$ \cite{Murthy:2009dq}. For other fillings we have $i \phi^0=c\tau+d$. With this in mind, we see from (\ref{ident global AdS3}) that $\psi+i t\sim \psi+i t +2\pi n +2\pi i m/\phi^0$ while $\psi-it\sim \psi-it+2\pi n $, that is, a torus with $\bar{\tau}'=-1/{\bar{\tau}}=0$ and $\tau'=-1/{\tau}$.

In the physical theory we have three dimensional supergravity coupled to $SU(2)_L$ and $SU(2)_R$ Chern-Simons terms. Following \cite{David:1999nr}, the precise content is
\begin{eqnarray}\label{3d sugra w CS}
S=&&\int d^3x\left[\sqrt{g}\left(R+2m^2\right)-\frac{1}{m}\epsilon^{\mu\nu\rho}\bar{\psi}_{\mu}D_{\nu}\psi_{\rho}\right]\nonumber\\
&&-\frac{k_R}{4\pi}\int \text{Tr}\left(A_R\wedge dA_R+2/3A_R^3\right)+\frac{k_L}{4\pi}\int \text{Tr}\left(A_L\wedge dA_L+2/3A_L^3\right)
\end{eqnarray}
The gauge connections $A_L,\,A_R$ correspond respectively to the $SU(2)_L$ and $SU(2)_R$ Chern-Simons terms and the field $\psi^{i}_{\mu}$ is a Dirac gravitino transforming in the fundamental of $SU(2)_R$ with index $i$. The covariant derivative is defined as $D_{\nu}=\partial_{\nu}+\omega_{ab\nu}\gamma^{ab}/4-m\gamma_{\nu}+A_{R}$, and the trace is taken in the fundamental representation. Because of supersymmetry the right Chern-Simons level is related to the cosmological constant as $k_R=4\pi/m$, while the left level $k_L$ is independent.

It is well known that we can write the Einstein-Hilbert term with negative cosmological constant as a pair of $SL(2,\mathbb{R})$ Chern-Simons terms with equal levels \cite{Achucarro:1987vz,Witten:2007kt}. The supergravity action (\ref{3d sugra w CS}) becomes a Chern-Simons action based on the supergroup $SU(1,1|2)_R\times SU(1,1)_L\times SU(2)_L$. In this case the gravitino transforms only under the fundamental of $SU(2)_R$ and thus the action splits into a non-supersymmetric (left) and a  supersymmetric (right) Chern-Simons actions, that is, 
\begin{equation}\label{CS 3d action}
S=S_L+S_R,
\end{equation}
with the non-supersymmetric $S_L$ action given by
\begin{equation}\label{S_L}
S_L=-\frac{ik_L}{4\pi}\int_{\mathcal{M}}\text{Tr}\left(\tilde{A}_L\wedge d\tilde{A}_L+\frac{2}{3}\tilde{A}_L^3\right)+\frac{ik_L}{4\pi}\int_{\mathcal{M}}\text{Tr}\left(A_L\wedge dA_L+\frac{2}{3}A_L^3\right)
\end{equation}
whereas the supersymmetric $S_R$ action is
\begin{eqnarray}
S_R=&&\frac{ik_R}{4\pi}\int_{\mathcal{M}}\text{Tr}\left(\tilde{A}_R\wedge d\tilde{A}_R+\frac{2}{3}\tilde{A}_R^3\right)-\frac{ik_R}{4\pi}\int_{\mathcal{M}}\text{Tr}\left(A_R\wedge dA_R+\frac{2}{3}A_R^3\right)\\
&&-\frac{ik_R}{4\pi}\int \bar{\psi}\wedge (d+\tilde{A}_R+A_R)\psi
\end{eqnarray}
Here $\tilde{A}_{L,R}$ denote respectively the $SL(2,\mathbb{R})_{L,R}$ connections. 

So far we have considered the diffeomorphic theory which corresponds to having equal left and right $SL(2,\mathbb{R})$ levels. Nevertheless, if the theory has gravitational Chern-Simons terms, then the levels can differ by  an amount proportional to the coefficient of those terms \cite{Hansen:2006wu}. On the other hand, the $SL(2,\mathbb{R})_R$ and $SU(2)_R$ levels must be the same because of supersymmetry. The $SU(2)_{L,R}$ levels in general have independent values. However, in the absence of gravitational Chern-Simons they are the same, so to preserve the full rotational symmetry- if we want to see this theory as coming from a truncation of six dimensional supergravity on a three sphere.

To fix the different levels as functions of the charges, we consider five dimensional supergravity reduced on $S^2$. In particular we are interested on the reduction of  five dimensional Einstein-Hilbert and abelian Chern-Simons terms. This was partially analyzed in \cite{Dabholkar:2014ema} and we review it now. The Einsten-Hilbert term has the form
\begin{equation}
\int d^5x\sqrt{g}\;c_{IJK}\sigma^I\sigma^J\sigma^k\,R
\end{equation} 
where $\sigma^I$ is the vector-multiplet scalar and $c_{IJK}$ has values $c_{1ab}=c_{a1b}=c_{ab1}=C_{ab}$. At the on-shell level we have the condition $\vartheta^{1/2}\sigma^I=p^I$. Thus after reduction on the sphere we obtain
\begin{equation}
\int d^3x\sqrt{g}\vartheta^{-1/2}\;p^1P^2\,R+\ldots.
\end{equation}
with $P^2=C_{ab}p^ap^b$. To go from the Einstein-Hilbert action to the Chern-Simons formulation, the metric must have unit constant curvature and therefore we scale it further  as 
\begin{equation}\label{metric rescaling}
 g_{\mu\nu}=\vartheta g^{(0)}_{\mu\nu}
\end{equation}
where $g_{\mu\nu}$ is asymptotically (\ref{AdS2xS1 metric}), to obtain 
\begin{equation}\label{EH normalized curv}
\int d^3x\sqrt{g^{(0)}}\;p^1P^2\,R^{(0)}.
\end{equation}
The Ricci curvature $R^{(0)}$ is now normalized to one at the on-shell level. It is the rescaled Einstein-Hilbert term (\ref{EH normalized curv}) that determines the Chern-Simons levels. These can be determined as in \cite{Hansen:2006wu} and hence we find 
\begin{equation}
 k_L=k_R=p^1P^2/2
\end{equation} 

 On the other hand the five dimensional Chern-Simons reduces to
\begin{eqnarray}\label{5d CS reduction}
&&\frac{\pi i}{3(4\pi)^3}\int_{S^2} c_{IJK}A^I\wedge F^J\wedge F^K\rightarrow \nonumber\\
&&-\frac{\pi i}{p^1(4\pi)^2}P^2\int A^1\wedge F^1
+\frac{\pi i p^1}{(4\pi)^2}\int C_{ab}\left(A^b+\frac{p^b}{p^1}A^1\right)\wedge\left(F^a+\frac{p^a}{p^1}F^1\right).
\end{eqnarray}
where $a=2\ldots N_V+1$, with $N_V$ the number of $\CN=4$ vector-multiplets in the four dimensional theory.
We have diagonalized the different couplings in order to identify the different Chern-Simons terms. The first term in (\ref{5d CS reduction}) can be interpred as coming from a $U(1)$ truncation of an $SU(2)_L$ Chern-Simons term. By writing 
\begin{equation}\label{SU(2) to U(1)}
 A_L=i\sigma^3/(2p^1)A^1
\end{equation}
the $SU(2)_L$ Chern-Simons action (\ref{S_L}) leads precisely to this $U(1)$ term.  Note that the factor of $1/p^1$ in (\ref{SU(2) to U(1)}) is necessary to have the correct Chern-Simons level $k_L=p^1P^2/2$, which in turn equals $k_R$. If we had considered instead  the six dimensional theory reduced on a three sphere we would have obtained directly a $SU(2)_L$ Chern-Simons action by gauging the left isometries of the three sphere and likewise an $SU(2)_R$ Chern-Simons from gauging the right isometries. 

Before moving to the one-loop computation we need to study the classical solutions of the Chern-Simons action (\ref{CS 3d action}). These correspond to flat connections and covariantly constant spinors, that is, 
\begin{eqnarray}
&&d\tilde{A}_{L,R}+\tilde{A}_{L,R}\wedge \tilde{A}_{L,R}=0,\;dA_{L,R}+A_{L,R}\wedge A_{L,R}=0,\\
&&(d+\tilde{A}_R+A_R)\psi=0.\label{eq gravitino}
\end{eqnarray}
The holonomies of these flat connections were reviewed in \cite{Dabholkar:2014ema}. On a general $SL(2,\mathbb{Z})$ filling, the Wilson lines are \footnote{The holonomies are defined up to conjugation.},
  \bea
\oint_{C_{n}}\tilde A_L =  2\pi i  \frac{a\tau+b}{c\tau+d} \frac{\sigma^3}{2} & \, ,& \qquad \oint_{C_{n}}\tilde A_R =  - 2\pi i  \frac{a}{c} \frac{\sigma^3}{2}\, ; \\
\oint_{C_{c}}\tilde A_L =  2\pi i   \frac{\sigma^3}{2} & \, , & \qquad \oint_{C_{c}}\tilde A_R =  - 2\pi i   \frac{\sigma^3}{2},\,\qquad ad-bc=1
\eea
and similarly for the $SU(2)$ connections. Here $C_n$ is the non-contractible cycle and $C_c$ is the contractible cycle. In terms of the coordinates $\theta,\,y$ in (\ref{AdS2xS1 metric}) we have $C_n=aC_1+bC_2$ and $C_c=cC_1+dC_2$ with $C_1=-\theta$ and $C_2=y$. 

Note that the holonomy of $\tilde{A}_R$ along the contractible cycle is minus one. This makes the gravitino antiperiodic; it is a well known fact that covariantly constant spinors are antiperiodic along contractible cycles. However, since the gravitino also couples to $A_R$ (\ref{eq gravitino}), which  has holonomy minus one too, it becomes effectively periodic. This is in agreement with the boundary conditions of the R-sector of the dual  CFT \cite{Dijkgraaf:2000fq}.

We are now ready to compute the modular anomaly using the super Chern-Simons path integral. To illustrate the main idea we consider first the case of Chern-Simons theory on a compact manifold. The non-compact case will follow by a straightforward generalization.

It is a well known fact that Chern-Simons theory with simple Lie group on a compact manifold is a topological theory \cite{Witten:1988hf}. This fact seems apparent from the Chern-Simons functional because it does not depend explicitly on a metric. However, at the quantum level we need to pick a metric  to ensure a well defined gauge fixed path integral \cite{Witten:1988hf} and thus a priori it is not obvious that this choice is not anomalous. The problem can arise from the appearance of zero modes which signal that the choice of metric is not compatible with the regulator used. As a matter of fact, studies of the exact Chern-Simons partition function on a manifold $M$   reveal that there can be a dependence on the volume at one-loop level. In particular, at large level $r$ we find \cite{Lisa92}
\begin{equation}\label{CS semiclassical}
Z(M)\simeq\sum_{A}e^{2\pi i r CS(A)}\tau(M,A)^{1/2}\;r^{(\text{dim}H^1_A-\text{dim}H^0_A)/2}\;\text{Vol}(M)^{(\text{dim}H^1_A-\text{dim}H^0_A)/2}
\end{equation}
The sum is over gauge equivalence classes of flat connections $A$. Both  the Chern-Simons invariant $CS(A)$ \cite{kirk1990} and $\tau(M,A)^{1/2}$, the Reidemeister-Ray-Singer torsion, are topological invariants. However we see that there can be a metric dependence via the term $\text{Vol}(M)^{(\text{dim}H^1_A-\text{dim}H^0_A)/2}$, where $\text{Vol}(M)$ is the volume of the manifold.

The metric anomaly can be explained succinctly as follows. In the gauge fixed theory we introduce an auxiliary bosonic scalar $b$ and ghost fermions $c,\bar{c}$ together with a metric to impose a Lorenz gauge fixing condition $d_A\star B=0$ with $B$ the gauge field fluctuation \cite{Witten:1988hf}. In computing the one-loop contribution there can be zero modes for the one-forms and scalars if the cohomology groups of the flat bundle, respectively  $H^1_A$ and  $H^0_A$, are not zero. Correspondingly the number of zero modes is
\begin{equation}\label{zero modes CS}
 N_{\text{zero}}^{0}=\text{dim}(H^0_A),\;N_{\text{zero}}^{1}=\text{dim}(H^1_A)
\end{equation}
for scalars and one-forms respectively.

Lets consider for example the case of the ghost zero modes $c,\bar{c}$. For the purpose of this work, it is suitable to take a manifold $M$ that results from a quotient $\tilde{M}/\Gamma$. In this case the zero modes correspond to the elements of the gauge group that commute with $\Gamma$, that is, the elements that leave the flat connection invariant. The zero modes sign a residual gauge symmetry and thus when we divide the path integral by the volume of the gauge group there is a factor of the volume of the residual gauge symmetry that remains in the denominator. To compute this volume we use the ultralocality argument explained in section \S\ref{vec measure sec}.
 In this case we consider a Grassmannian measure for the gauge fixing ghosts of the form
\begin{equation}\label{CS normalization grass}
\int [d\bar{c}][dc]\exp{\left\{-r\int \text{Tr} \,\bar{c}\wedge\star c\right\}}=1.
\end{equation}
The operator $\star$ is defined with respect to a metric with constant curvature $\pm1$ on $M$. By choosing a basis of orthonormal adjoint-valued  eigenfunctions of the Laplacian on $\tilde{M}$, the normalization (\ref{CS normalization grass}) leads to the measure
\begin{equation}
[d\bar{c}][dc]=\prod_{x,\mu}\left(\frac{r}{|\Gamma|}\right)^{-1}d\bar{c}\,dc
\end{equation}
by the usual rules of grassmann integration. Integrating this measure over the space of zero modes we obtain
\begin{equation}
 \left(\frac{r}{|\Gamma|}\right)^{-\text{dim}(H^0_{A})}
\end{equation}
where $\text{dim}(H^0_{A})$ is the number of ghost zero modes (\ref{zero modes CS}). Similarly, from the scalar bosonic zero modes $b$ we find a contribution of $(r/|\Gamma|)^{\text{dim}(H^0_{A})/2}$. In this case the exponent is just $1/2$ times the dimension of $H^0_{A}$ because we are dealing with  real bosonic scalars. Therefore, the total zero mode contribution from both bosonic and fermionic scalars is $(r/|\Gamma|)^{-\text{dim}\,H^0_A(M)/2}$.

A similarly exercise for one-forms, using the normalization
\begin{equation}\label{non-abelian ultralocality}
\int [dA]\exp{\left\{-r\int \text{Tr} \,A\wedge\star A\right\}}=1,
\end{equation} gives the zero mode contribution
\begin{equation}\label{1-form zero contrib}
 \left(\frac{r}{|\Gamma|}\right)^{\text{dim}H^1_A(M)/2}.
\end{equation}
Equivalently we can replace $|\Gamma|$ by the volume of the manifold since we have $\text{vol}(M)=\text{vol}(\tilde{M})/|\Gamma|$. This is particularly adequate when the group $\Gamma$ has infinite order, as it is the case of the quotient $AdS_2\times S^1=AdS_3/\Gamma$, which is our main interest. Putting together  the one-form and the scalar zero mode contributions we obtain the metric dependence of the one-loop determinant (\ref{CS semiclassical}), including the exact dependence on the Chern-Simons level $r$.

With this in mind we turn gears to the non-compact case.  The key idea is to look at $AdS_2\times S^1$ as the quotient of $AdS_3$ by the additive group $\Gamma$ (\ref{ident global AdS3}). We compute the $AdS_3$ partition function using microcanonical boundary conditions. These consist in fixing the Wilson lines along the boundary cycle $C_2$ parametrized by the coordinate $y$ in (\ref{AdS2xS1 metric}) and summing over the Wilson lines along the cycle $C_1$ parametrized by $\theta$ \cite{Dabholkar:2014ema}. For large level $r$, the partition function has the form
\begin{equation}\label{AdS3 CS 1-loop}
Z_{\text{AdS}_3|\text{micro}}\simeq\sum_{A}e^{2\pi i r \text{CS}(A)}Z_{1\text{-loop}}
\end{equation}
where $A$ are flat connections on $AdS_2\times S^1$ and $\text{CS}(A)$ is the Chern-Simons action of the flat connection, properly renormalized by boundary counter terms. The Chern-Simons action of these flat connections was studied in \cite{Dabholkar:2014ema} for example.

The component $Z_{1\text{-loop}}$ arises from the determinant over the non-zero modes. This determinant consists of two different contributions: one is a local contribution coming from a zeta function regularized determinant which is topological \cite{Carlip:1992wg,Bytsenko:1998uy}, and so it does not depend on either $r$ or $|\Gamma|$. The other contribution comes from the zero modes, which we now describe. 

On $AdS_2\times S^1$  we need to care only about the one-form zero modes because constant functions are non-normalizable. In particular, the zero modes correspond to adjoint-valued connections that are closed under $d+A$ with $A$ the flat connection. In this problem the flat connection is never the trivial solution $A=0$, and thus the zero modes correspond to  adjoint-valued closed forms  that commute with the flat connection. For each factor of the gauge group the maximal commuting subgroup is always one-dimensional and therefore we only need to find the space of closed one-forms modulo gauge transformations in the De Rham sense. 

Since on $AdS_2\times S^1$ we fix the Wilson lines along the circle $S^1$, because of the microcanonical boundary conditions, we can discard the flat connections of the type $\sim dy$. Therefore the only zero modes we can have are those along the $AdS_2$ directions, which are precisely the ones we have determined in section \S \ref{vec measure sec} equation (\ref{U1 zero mode}). The measure for the zero modes is determined analogously to the compact case except that here the order of $\Gamma$ is infinite.  The way to proceed is to define $|\Gamma|$ from its action on the volume of the quotient manifold and so we have effectively
\begin{equation}
 |\Gamma|=\phi^0
\end{equation}
where $1/\phi^{0}$ is the radius of the circle $S^1$ in the metric (\ref{AdS2xS1 metric}). In addition, we need to regularize the infinite number of zero modes by introducing boundary counter terms as we did in section \S\ref{vec measure sec}. The zero mode contribution is then
\begin{equation}
 \left(\frac{k}{|\Gamma|}\right)^{\text{Ren}(N_{\text{zero}})/2}=\left(\frac{k}{|\Gamma|}\right)^{-1/2}
\end{equation}
with $k$ the corresponding Chern-Simons level. We have used the fact that the renormalized number of zero modes is $\text{Ren}(N_{\text{zero}})=-1$.

This is not the final answer because the gravitino can also have zero modes. On the background solution we can set $\psi_{\mu}=\bar{\psi}_{\mu}=0$, so the equation for the fluctuation zero mode $\delta\psi_{\text{zero}}$ is 
\begin{equation}
(d+\tilde{A}^{\text{flat}}_{R}+A^{\text{flat}}_R)\delta\psi_{\text{zero}}=0
\end{equation} 
where $\tilde{A}^{\text{flat}}_{R},A^{\text{flat}}_R$ are the on-shell $SL(2)_R$ and $SU(2)_R$ flat connections respectively, which as we have shown carry non-trivial holonomies. Since these are flat connections we can write the above equation as 
\begin{equation}
(d+g^{-1}dg)\delta\psi_{\text{zero}}=0 \Leftrightarrow d(g\delta\psi_{\text{zero}})=0
\end{equation}
with $g$ a gauge transformation element in $SL(2)_R\times SU(2)_R$.
Much like for the gauge connections, the gravitino zero modes therefore correspond to normalizable solutions of the form $g^{-1}d\epsilon$ with non-normalizable fermionic parameters $\epsilon$ \cite{Banerjee:2011jp}- these are the solutions that can not be gauged away by a normalizable gauge transformation.  Besides, there are auxiliary fermionic scalars and bosonic scalar ghosts in the gauge fixed theory  but, as explained, they do not lead to additional zero modes because a constant scalar is not normalizable on $AdS_2\times S^1$. To compute the measure for the gravitino zero modes we use the $SU(2)_R$ invariant measure
\begin{equation}\label{gravitino measure def}
 \int[D\psi] [D\bar{\psi}]\exp{\left\{-r\int \bar{\psi}\wedge\star \psi\right\}}=1
\end{equation}
where $\psi$ is the gravitino one-form that transforms in the fundamental of  $SU(2)_R$. The ultralocality argument gives the zero-mode volume
\begin{equation}
 \left(\frac{r}{|\Gamma|}\right)^{-\tilde{N}_{\text{zero}}}
\end{equation}
where $\tilde{N}_{\text{zero}}$ is the number of gravitino one-form zero modes. This number is infinite so we have to proceed as in section \S\ref{vec measure sec}. The holonomies of the flat connections $\tilde{A}^{\text{flat}}_{R},A^{\text{flat}}_R$ have residual gauge symmetries that we can use to bring $\delta\psi$ to a particular direction in the $SL(2)_R\times SU(2)_R$ space. In particular, the flat connection is invariant under $g\rightarrow h g$ with $h$ a constant element of $SL(2)_R\times SU(2)_R$. This leads to the transformation $d\epsilon \rightarrow h^{-1}d\epsilon$. Therefore the zero-mode  effectively consists of a complex Grassmann-valued one form expanded in the basis (\ref{U1 zero mode}). We regularize the number of zero-modes as
\begin{equation}
 \tilde{N}_{\text{zero}}=\sum_{l}\langle\psi_l|\psi_l\rangle
\end{equation}
where $l$ is the quantum number that parametrizes the zero mode and $\langle\;|\;\rangle$ is an $SU(2)_R$ invariant norm induced from the measure (\ref{gravitino measure def}). Much like for the gauge connections we find
\begin{equation}
 \tilde{N}_{\text{zero}}=\sum_{l}\langle\psi_l|\psi_l\rangle=\cosh(\eta_0)-1
\end{equation}
where $\eta_0$ is an $AdS_2$ cuttoff. The renormalized number of zero modes is therefore 
\begin{equation}
 \tilde{N}_{\text{zero}}^{\text{ren}}=-1,
\end{equation}
and so from the gravitino zero modes we obtain the contribution $r/|\Gamma|$. 

We are now ready to assemble the different contributions. From the supersymmetric side we have bosonic zero modes for both the $SL(2,\mathbb{R})_R$ and $SU(2)_R$ connections which give a total contribution of $\sim |\Gamma|/k_R$, as they have equal Chern-Simons levels. This cancels the gravitino zero mode contribution $\sim k_R/|\Gamma|$. Thus the total contribution from the supersymmetric side is trivial
\begin{equation}
 Z_{1\text{-loop }S_R}\sim 1
\end{equation}
On the other hand, from the non-supersymmetric side we have the contribution from only the gauge connections zero modes which give
\begin{equation}\label{1-loop result}
 Z_{1\text{-loop }S_L}\sim \frac{|\Gamma|}{\sqrt{\tilde{k}_Lk_L}}
\end{equation}
where $\tilde{k}_L$ is the $SL(2)_L$ level and $k_L$ is the $SU(2)_L$ level. In terms of the cohomology groups $\hat{H}^{0,1}$ defined in section \S\ref{vec measure sec} for $AdS_2$, the one-loop contribution (\ref{1-loop result}) has precisely the same form as the compact result (\ref{CS semiclassical}). Note that, even though we are taking the level $k$ large to compute the one-loop contribution, the result (\ref{1-loop result}) is expected to hold even for small $k$ since it relies purely on a zero mode argument that depends only on the cohomology structure of the space. 

In the same way, the contribution from the abelian gauge fields is
\begin{equation}
 Z_{1\text{-loop}}^{U(1)}=\left(\frac{p_1}{|\Gamma|}\right)^{-N_{V}/2}
\end{equation}
where $p_1$ is the $U(1)$ Chern-Simons level (\ref{5d CS reduction}) and $N_V$ is the number of $\CN=4$ vector-multiplets. Note that this agrees with the vector-multiplet computation of section \S\ref{vec measure sec}.

There is however an additional contribution besides the one-loop Chern-Simons correction. As explained, in the Chern-Simons formulation we use a background metric with constant unit curvature. So in going from the physical theory (\ref{3d sugra w CS}) to the Chern-Simons formulation there is a rescaling of the metric (\ref{metric rescaling}). This leads to an additional contribution to the $SL(2)_{L,R}$ Chern-Simons measure as function of $\vartheta$, the size of the physical metric. That is, the measure for the physical metric is effectively
\begin{equation}
 [Dg]\sim [\vartheta^{1/2} D\tilde{A}_L][ \vartheta^{1/2}D\tilde{A}_R]
\end{equation}
where $\vartheta$ is the size of the on-shell metric, which is a constant. By the same zero mode counting argument we find an additional contribution proportional to $\vartheta$ from integrating the $SL(2)$ gauge fields. To be more explicit, define $A^g_{L,R}\equiv \vartheta^{1/2}\tilde{A}_{L,R}$. The ultralocality normalization (\ref{non-abelian ultralocality}) now becomes 
\begin{equation}
\int [dA^g]\exp{\left\{-\frac{r}{\vartheta}\int \text{Tr} \,A^g\wedge\star A^g\right\}}=1
\end{equation}
The exponential factor is just the norm of the Chern-Simons variables $\tilde{A}_{L,R}$. The measure $[dA^g]$ must carry a factor of $(r/\vartheta|\Gamma|)^{1/2}$ for each mode. So integrating over the zero mode space we obtain $\vartheta^{-N_{\text{zero}}/2}=\vartheta^{1/2}$ for each gauge group factor. We remind the reader that in the path integral we are not really integrating over the zero modes as they give rise to infinities. Instead the ultralocality normalization is a nice way to correctly regulate the integration over the non-zero modes. To put in other words, we are instructed to compute a one-loop correction from the gravitational path integral, which we can rewrite in terms of the Chern-Simons variables, that is, 
\begin{equation}
\int [Dg] \exp{\left[\int R-2\Lambda+\ldots\right]}|_{1\text{-loop}}=\int \prod_{n,m\in\text{non-zero}} [\vartheta^{1/2}D\tilde{A}_L]_{n} [\vartheta^{1/2}D\tilde{A}_L]_{m} \exp{\left[\int CS(A_{L,R})\right]}|_{1\text{-loop}}
\end{equation}
where $\Lambda$ is the cosmological constant and $R$ is the Ricci scalar. From this point of view, the measure has an additional $\vartheta^{1/2}$ factor for each non-zero mode.  So integration over these gives $\vartheta^{N_{\text{non-zero}}/2}$ which equals $\vartheta^{(N_{\text{total}}-N_{\text{zero}})/2}$. The term with $\vartheta^{N_{\text{total}}/2}$ can be renormalized to zero because it is an ultra-local function, and hence we obtain $\vartheta^{-N_{\text{zero}}/2}=\vartheta^{1/2}$, which is consistent with the ultralocality argument.

The total contribution from the $SU(1,1|2)\times SL(2)_L\times SU(2)_L$ and abelian Chern-Simons is therefore
\begin{equation}\label{CS 1-loop final}
 Z_{1\text{-loop}}\sim \vartheta\frac{|\Gamma|}{\sqrt{\tilde{k}_Lk_L}}\left(\frac{|\Gamma|}{p^1}\right)^{N_V/2}
\end{equation}
Since this contribution leads to a correction of the effective action of the form $\sim \ln|\Gamma|$ it cannot be renormalized by a local counterterm and therefore represents an anomaly.

We can now compare the Chern-Simons one-loop correction with the one-loop approximation of the localization integral  on $AdS_2\times S^2$ (\ref{loc partition fnct}). Introducing a measure $\mathcal{M}(\phi^0)$ we have
\begin{equation}\label{1-loop loc}
 Z_{AdS_2,\,1\text{-loop}}\simeq e^{\frac{\pi}{2}\frac{\Delta}{p^1P^2}\phi^0+\frac{\pi}{2}\frac{p^1}{\phi^0}(P^2+8c_1)}\times \mathcal{M}(\phi^0)\frac{(\phi^0)^2}{\sqrt{P^2(P^2+8c_1)}}\left(\frac{\phi^0}{p^1}\right)^{N_V/2}
\end{equation}
Here we have used the zero instanton prepotential (\ref{zero prepotential}); the instanton sector gives rise to exponentially suppressed corrections. In this expansion, it is enough to scale $\Delta\gg 1$ for fixed but large $P^2$. This is so because we can write the exponential in the form $\sim \Delta^{1/2}(x+1/x)$, with $p^1/\phi^0=x\sqrt{Q^2/P^2}$, and we have approximated $P^2+8c_1\simeq P^2$; we are assuming the measure is at most polynomial so it does not change the on-shell attractor values. Therefore, in the saddle approximation we are expanding in powers of $\Delta^{1/2}\gg 1$. At the saddle, $x\sim 1$ and so we can consider arbitrary values of $Q^2$ while keeping $\Delta\sim Q^2P^2\gg 1$ for $P^2\gg 1$. This means that in the saddle approximation we can have arbitrary on-shell values of $\phi^0\sim \sqrt{Q^2/P^2}$ by dialing the value of $Q^2$. 

In the one-loop approximation (\ref{1-loop loc}) the integrals over $\phi^0$ and $\phi^1$ give respectively the terms $(\phi^0)^{3/2}/\sqrt{(P^2+8c_1)}$ and $(\phi^0)^{1/2}/\sqrt{P^2}$, while the term $(\phi^0/p^1)^{N_V/2}$ arises from integrating over the $\phi^a$, that is, the vector-multiplet integrals. Furthermore, the exponential term can be identified with the Chern-Simons integral for the flat connections \cite{Dabholkar:2014ema}. That is, the on-shell contribution from  the supersymmetric Chern-Simons gives exactly zero, while on the non-supersymmetric side there are two types of contributions: there is a boundary contribution of $\Delta\phi^0/p^1P^2$, due to the microcanonical boundary conditions, and a term $p^1(P^2+8c_1)/\phi^0$  which comes from the bulk integral. In fact, from the bulk term we can identify the $SL(2,\mathbb{R})_L$ level as 
\begin{equation}
 \tilde{k}_L=p^1(P^2+8c_1)/2 
\end{equation}
which agrees with other computations in higher dimensional gravity \cite{Kraus:2005vz,Banerjee:2011ts}. Therefore, at the on-shell level we can match the renormalized action $\frac{\pi}{2}\frac{\Delta}{p^1P^2}\phi^0+\frac{\pi}{2}\frac{p^1}{\phi^0}(P^2+8c_1)$, with the Chern-Simons action of the $SL(2)_L$ flat connection \cite{Dabholkar:2014ema}.

Note that the Chern-Simons one-loop computation (\ref{CS 1-loop final}) holds for any value of $|\Gamma|=\phi^{0}$, since it corresponds to a different choice of background metric. Similarly, the one-loop approximation (\ref{1-loop loc}) is valid for arbitrary values of $\phi^0$. Equality of the one-loop term in (\ref{1-loop loc}) with (\ref{CS 1-loop final})  determines the measure 
\begin{equation}\label{measure final}
 \mathcal{M}(\phi^0)\sim \frac{\vartheta}{\phi^0 p^1}.
\end{equation}
 This is our main result. 

We can compute the charge dependence of the $\vartheta$ from the zero instanton prepotential. Using  the formulas (\ref{size high der corrct}) and (\ref{size inst corrections}) we compute
\begin{equation}
 \vartheta=P^2+4c_1.
\end{equation}
Together with (\ref{measure final}) we obtain the measure for one-quarter BPS black holes
\begin{equation}\label{measure 1/4}
 \mathcal{M}_{1/4}(\phi^0)=\frac{P^2+4c_1}{\phi^0p^1}
\end{equation}
This measure reproduces the exact leading Bessel function, including the value of the polar coefficient, in the microscopic answer (\ref{finite sum Bessels}) for the $\CN=4$ CHL models in both $K3$ and $T^4$. 

For the one-eighth BPS case, we have exactly $\vartheta=P^2$ and so the measure is
\begin{equation}\label{measure 1/8}
 \mathcal{M}_{1/8}(\phi^0)=\frac{P^2}{\phi^0p^1}.
\end{equation}
Together with the contribution $Z_{\text{odd}}$ (\ref{Zodd}) coming from the odd fields, we obtain an exact agreement with the microscopic answer (\ref{deg 1/8 integral}). 

Finally, note that for large $P^2$ the measures (\ref{measure 1/4}) and (\ref{measure 1/8}) are asymptotically the same. This is consistent with the fact that they arise from integrating out the fields in the $\CN=4$ supergravity multiplet. Since in that limit of charges, the attractor background is exactly the same in both cases,  we expect to obtain the same contribution to the measure after integrating out the massless fields in the $\CN=4$ supergravity multiplet. 
On the other hand, the factor $1/\phi^0p^1$ is universal. By combining it with $d\phi^0d\phi^1$ it ensures the $SL(2,\mathbb{R})$ invariant measure $d\tau d\bar{\tau}/(\tau_2)^2$ where $\tau=\phi^1/\phi^0+ip^1/\phi^0$ is the complex structure of the torus,  which is part of the six dimensional metric $AdS_2\times S^2\times T^2$. This is as expected since from a six dimensional point of view we can either reduce the theory to four dimensions by first going down on one circle and then on the other or on any $SL(2,\mathbb{Z})$ combination of these.

% To compute $\text{Vol}(G_0)$, we introduce the path integral measure \cite{Carlip:1992wg,Banerjee:2010qc}
% \begin{equation}\label{CS normalization}
% \int [d\omega]\exp{\left\{-r\int \text{Tr} \,\omega\wedge\star \omega\right\}}=1
% \end{equation}
% where $\omega$ is a bosonic zero-form and  the operator $\star$ is defined with respect to a metric with constant curvature $\pm1$ on $M$. If we choose a basis of orthonormal adjoint valued  eigenfunctions of the Laplacian on $\tilde{M}$, then the normalization (\ref{CS normalization}) leads to the measure
% \begin{equation}
% [d\omega]=\prod_{x,\mu}\left(\frac{r}{|\Gamma|}\right)^{1/2}d\omega(x)
% \end{equation}
% where $|\Gamma|$ is the order of the group $\Gamma$.
% Integrating this measure over the zero modes we obtain the volume $\text{Vol}(G_0)$ of the residual gauge symmetry 
% \begin{equation}\label{vol G0}
% \text{Vol}(G_0)\propto \left(\frac{r}{|\Gamma|}\right)^{\text{dim}(G_0)/2}.
% \end{equation}

\subsection{Instanton corrections}\label{sec CS instantons}

In section \S\ref{micro N=4 sec} we found a formula for the degeneracy of one-quarter BPS states which contains subleading Bessel corrections (\ref{finite sum Bessels}). In this section we explore the contribution of instantons  in the localization integral (\ref{loc partition fnct}) and argue that they are responsible for the subleading Bessel functions. Nonetheless, we do not explain how these instantons are included at the level of the path integral. Instead we use an effective description in terms of an instanton quantum corrected prepotential. We tailor this answer in such a way that it is easy to read the effect of the instantons and interpret this in terms of an effective Chern-Simons theory that we can work with.  

Lets start with the effective $\CN=2$ non-perturbative  prepotential $F^{\text{non-pt}}(X)$. This was computed in  \cite{Harvey:1996ir} and can be obtained as the holomorphic part of an $R^2$ amplitude at one-loop. For the $\CN=4$ theories the prepotential is one-loop exact and has the form
\begin{equation}\label{non-pt prepotential}
F^{\text{non-pt}}=-\frac{1}{2}\frac{X^1}{X^0}C_{ab}X^aX^b-W^2\ln g\left(\frac{X^1}{X^0}\right)
\end{equation}
where $W^2$ is the square of the on-shell value of the graviphoton field. The function $g(\tau)$ is a modular form of a congruence subgroup and has precisely the form given in (\ref{K3 g(t)}) and (\ref{T4 g(t)}) for respectively the $K3$ and $T^4$ orbifold compactifications. From the heterotic point of view, the contributions coming from $g(\tau)$ can be interpreted as NS-5 brane instantons corrections to the tree level $R^2$ amplitude  \cite{Harvey:1996ir}. 

As it is well known, the $\CN=2$ prepotential encodes a series of $R^2$ corrections in the low energy $\CN=2$ supergravity action of the form  
\begin{equation}\label{prepotential high derivatives}
\sim \frac{\partial F}{\partial W^2}\,R^{-}\wedge R^{-}+c.c.
\end{equation}
where $R^{-}$ is the anti-self-dual part of the Riemann tensor and $c.c.$ stands for complex conjugate. For the $AdS_2\times S^2$ geometry, the non-perturbative prepotential (\ref{non-pt prepotential}) encodes, via the terms (\ref{prepotential high derivatives}), the Gauss-Bonnet corrections discussed in section \S\ref{N=4 sugra sec}, that are known to exist in the Heterotic frame.

Introducing the quantum corrected prepotential (\ref{non-pt prepotential}) at the level of the Wilsonian action can be problematic, specially when we want to compute the path integral using localization. The reason is twofold. First, the function $\ln g(\tau)$ in (\ref{non-pt prepotential}) is generically singular at finite values of $\tau$. This can spoil the localization argument since in this case we go  off-shell. The second reason is that, from a five dimensional point of view, $\CN=2$ supersymmetry fixes the fourth derivative terms to be schematically of the form $\psi\, R\, R$, with $\psi$  some field and therefore these couplings cannot accommodate instantonic contributions. The only exception however, is the zero instanton contribution. In this case the reduction of five dimensional supergravity down to four dimensions leads to a theory with prepotential the zero instanton approximation of  (\ref{non-pt prepotential}) \cite{Banerjee:2011ts}, that is, 
\begin{equation}\label{zero inst prepotential}
F^{(0)}=-\frac{1}{2}\frac{X^1}{X^0}C_{ab}X^aX^b+c_1\,W^2\frac{X^1}{X^0}
\end{equation}
where $c_1$  takes the values $0,1$ for the $T^4$ and $K3$ orbifold compactifications respectively. For these reasons, we take the zero instanton approximation (\ref{zero inst prepotential}) as the microscopic prepotential. It would be interesting to understand the instanton corrections as coming from additional saddles in the localization computation along the lines of \cite{Beasley:2006us}.

The localization formula for the $AdS_2$ path integral (\ref{loc partition fnct}) is, nevertheless, generic for any holomorphic prepotential. This is based on the assumption that the localization  saddles (\ref{localization solution}) are not changed after taking into account the instantons. In this case, the renormalized action on the localization locus leads precisely to the integral (\ref{loc partition fnct}) with the non-perturbative prepotential. 

Lets proceed with this assumption and continue by studying the integral (\ref{loc partition fnct}) with the non-perturbative prepotential. After all, it reproduces the microscopic answer (\ref{OSV 1/4}) up to a measure factor. We obtain
\begin{equation}
 d(q,p)\sim \int d\mu(\tau,\bar{\tau})\exp\left[\frac{\pi}{2}\frac{|Q+\tau P|^2}{\tau_2}-\Omega(\tau,\bar{\tau})\right]
\end{equation}
with
\begin{equation}
 \Omega(\tau,\bar{\tau})=\ln g(\tau)+\ln g(-\bar{\tau})
\end{equation}
and  $\tau=\tau_1+i\tau_2=\phi^1/\phi^0+ip^1/\phi^0$ and $\bar{\tau}=\tau_1-i\tau_2$. We have denoted the measure by $d\mu(\tau,\bar{\tau})$. Developing on this formula, we expand the exponential $\Omega(\tau,\bar{\tau})$ in Fourier series
to obtain
\begin{eqnarray}\label{npt localization expnd}
 d(q,p)&&\sim \sum_{n_1,n_2=0}^{\infty}\int d\mu(\tau,\bar{\tau};n_1,n_2)\,d(n_1)d(n_2)\,\exp\left[\frac{\pi}{2}\frac{|Q+\tau P|^2}{\tau_2}+4\pi\tau_2c_1+2\pi i\tau n_1-2\pi i\bar{\tau}n_2\right]\nonumber\\
{}
\end{eqnarray}
where we have used the fact that
\begin{equation}
 \exp{[-\Omega(\tau,\bar{\tau})]}=e^{4\pi\tau_2 c_1}\sum_{n_1,n_2=0}^{\infty}d(n_1)d(n_2)e^{2\pi i\tau n_1}e^{-2\pi i\bar{\tau}n_2}
\end{equation}
with $d(n)$  the Fourier coefficients of $1/g(\tau)$. 

Formula (\ref{npt localization expnd}) is suggestive of an instanton/anti-instanton sum. The exponential 
\begin{equation}\label{zero inst loc contribution}
 \frac{\pi}{2}\frac{|Q+\tau P|^2}{\tau_2}+4\pi\tau_2c_1
\end{equation}
is the answer that one obtains using the zero instanton prepotential (\ref{zero inst prepotential}), that is, the contribution coming from the supergravity fields. On the other hand, inspired by the analysis of \cite{Beasley:2006us}, the sum in (\ref{npt localization expnd}) can be interpreted as the contribution coming from  $d(n_1)$ worldsheet instantons at the north pole of $S^2$ and $d(n_2)$ anti-instantons at the south pole, with charges $n_1$ and $n_2$ respectively.
  
In each sector $(n_1,n_2)$ we can compute the effect of the instantons to the attractor background geometry. Namely, 
we can compute the effective $AdS_2$ size $\vartheta(P^2,n_1,n_2)$ with the formula (\ref{size high der corrct}) but now with respect to an effective entropy function defined by (\ref{npt localization expnd}):
\begin{equation}\label{effective entropy inst}
 \mathcal{E}_{n_1,n_2}=\frac{\pi}{2}\frac{|Q+\tau P|^2}{\tau_2}+4\pi\tau_2c_1 -2\pi\tau_2 (n_1+n_2)+2\pi i\tau_1(n_1-n_2)
\end{equation}
We assume that the last two terms do not generate any dependency on either $\vartheta_1$ or $\vartheta_2$, as they seem to arise from couplings to topological terms. Therefore, proceeding as in section \S\ref{N=4 sugra sec}, we obtain again the attractor equation (\ref{size attractor eq}). That is, we have $\vartheta_1=\vartheta_2=\vartheta$ with
\begin{equation}\label{effective ads size}
 \vartheta(P^2,n_1,n_2)=P^2+4c_1-2(n_1+n_2).
\end{equation}
Since $\vartheta$ must be positive, we see that we can introduce instantons and anti-instantons up to a maximum charge of 
\begin{equation}\label{bound 1}
 n_1+n_2<P^2/2+2c_1
\end{equation}
is saturated. For this reason we call (\ref{bound 1}) the unitary bound. Furthermore, at the on-shell level, the value of the dilaton $\tau_2$ must be real. From the extremization of $\mathcal{E}_{n,m}$ we find a further restriction given by the positivity condition
\begin{equation}
 P^2/2+4c_1-2(n_1+n_2)+2(n_1-n_2)^2/P^2>0.
\end{equation}
Together with the fact that at the on-shell level the coupling constants $\tau$  and $\bar{\tau}$ in (\ref{npt localization expnd}) must have respectively positive and negative imaginary parts; this is the condition that instantons are negligible  at weak coupling,  we must have in addition 
\begin{equation}
 -P^2/2\leq n_1-n_2\leq P^2/2.
\end{equation}

For this reason, the infinite sum in (\ref{npt localization expnd}) is physically  truncated to  
\begin{equation}\label{truncated}
 d(Q,P)\simeq \sum_{n=0}^{P^2/2+2c_1}\sum_{\substack{m=0\\ -P^2/2\leq n-2m\leq P^2/2\\ \mathcal{F}(n,m)>0}}^{n}d(n-m)d(m)e^{2\pi i\frac{Q.P}{P^2}(n-2m)}\,\mathcal{I}_{(n,m)}(Q^2,P^2,Q.P)
\end{equation}
with the integral $\mathcal{I}(n,m)$ defined as
  \begin{eqnarray}\label{I integral}
 \mathcal{I}(n,m)=&&\int_{\tilde{\mathcal{C}}} d\mu(\tau_2,\tau_1;n,m)\exp{\left[\frac{\pi}{2}\frac{\Delta/P^2}{\tau_2}+4\pi\tau_2\mathcal{F}(n,m;P^2)\right]}\exp{\left[\frac{\pi}{2}\frac{P^2}{\tau_2}(\tau_1-Q.P/P^2)^2\right]}.\nonumber \\
{}
\end{eqnarray}
We have redefined $\tau_1\equiv \tau_1+2i \tau_2(n-2m)/P^2$, and $\mathcal{F}(n,m;P^2)$ is the function (\ref{Besse arg}). The contour $\tilde{\mathcal{C}}$ takes $\tau_1$ over the imaginary axis $i\mathbb{R}$ and $\tau_2$ along $\epsilon+i\mathbb{R}$ with $\epsilon>0$. This choice, also justified in \cite{Dabholkar:2011ec},  is based on the fact that the localization equations do not impose other restrictions on the parameters $\phi^a$ (\ref{loc partition fnct}) and thus have to integrated over an indefinite and convergent contour. In particular, we have chosen $\tau_2$ to lie along the steepest descent contour which is the imaginary line at $\epsilon+i\mathbb{R}$ while for $\tau_1$ we have chosen it to run over the imaginary values.

From the truncated sum (\ref{truncated}), it becomes clear that there are subleading saddle points to the leading zero instanton contribution, which is the physical attractor background. These subleading saddle points carry different expectation values for the dilaton $\tau_2$ and the size $\vartheta$, for example. This may look puzzling from the $AdS_2$ path integral point of view, since we are instructed to sum over fields that respect the same boundary conditions. However, this ties well with the fact that we are summing over instanton constributions. It is well known that instanton contributions arise due to non-local field configurations that carry topological charge and hence, even though they do not change the local equations of motion and therefore the physical boundary conditions, they can contribute non-trivially to effective action. 

The integrand (\ref{I integral}) still has the form of a Bessel times a gaussian even after including the instanton contributions. As explained in section \S\ref{sec CS}, the gaussian is a consequence of the unique bosonic zero mode, after an IR renormalization, of the $SU(2)_L$ connection whereas the Bessel exponential comes from evaluating the supersymmetric $SL(2)$ Chern-Simons on the $AdS_2\times S^1$ solution with microcanonical boundary conditions. This suggests that the effect of introducing the instantons is to renormalize the different couplings in the Chern-Simons theory. Namely, the term in the exponential 
\begin{equation}
 4\pi\tau_2\mathcal{F}(n,m;P^2)=\frac{\pi}{2}\tau_2\left(P^2+8c_1-4n+4(n-2m)^2/P^2\right)
\end{equation}
suggests that the tree level $SL(2)_L$ Chern-Simons level $P^2/2+4c_1$ gets renormalized by a factor of $-2n+2(n-2m)^2/P^2$. At first sight, this looks unnatural because, in general, the Chern-Simons coupling is integrally quantized. It might be possible, nevertheless, that there are additional contributions, namely, in the form of Kloosterman sums that render the answer well-defined under large gauge transformations, which is the reason why the level is quantized.  On the other hand, the gaussian term in (\ref{I integral}) is still proportional to $P^2$ which ensures that the $SU(2)_L$ level is not renormalized. 

With this assumption, we compute the measure using the super Chern-Simons formulation explained in the last section. This leads to 
\begin{equation}\label{inst measure}
 d\mu(\tau,\bar{\tau},n)=d^2\tau\,\frac{\vartheta(P^2,n)}{(\tau_2)^{k+3}}=\frac{d^2\tau}{(\tau_2)^{k+3}}(P^2+4c_1-2n)
\end{equation}
in each instanton sector. Plugging this back in the integral $\mathcal{I}_{(n,m)}$, the truncated sum (\ref{truncated}) is then in perfect agreement with the microscopic answer (\ref{finite sum Bessels}).

\section{Conclusions and open problems}

With this work we came a step closer to understand in full detail the exact entropy of supersymmetric black holes. The dream is to be able to show the equation which is at the heart of this program \cite{Sen:2008vm}, that is,
\begin{equation}
 Z_{AdS_2}=d(q).
\end{equation}
In particular, we want to understand how quantum gravity can reproduce the precise integers $d(q)$. A priori, it looks puzzling how a gravity path integral can reproduce such a sensible quantity to UV dynamics like an integer. From a bulk point of view there is no reason  why  $Z_{AdS_2}$ has to fulfill such requirement. Instead, the reason stems purely from holography: the dual $CFT_1$ has an hamiltonian with only a finite number of ground states \footnote{We are assuming that the quantum mechanics has a mass gap. For the problem at hands this gap is small but non-zero \cite{Sen:2011cn}.} and thus from this side the partition function is an integer.         

With this in mind, our results show an interesting interplay between quantum gravity and number theory. To this end, the main results are
\begin{itemize}
 \item An exact formula for the $AdS_2$ path integral that encodes all the power law corrections to the black hole entropy area formula in both $\CN=8$ and $\CN=4$ compactifications. In essence, this formula is characterized by a Bessel like behaviour whose index is captured by an anomaly coefficient in Chern-Simons theory.
 
\item The existence of subleading saddle points coming from instanton contributions. Also in this case, we find subleading Bessel like corrections which are in perfect agreement with the microscopic formulas. Along with this, we find an unitary condition given by the positivity of the effective $AdS_2$ size (\ref{effective ads size}) which leads to a truncation of the instanton sum.

\end{itemize}

Despite this success, there are still a few questions  we need to understand. The most urgent is the study of the measure purely from an $\text{AdS}_2$ point of view. As we mentioned before, this is a very difficult problem which requires understanding equivariant cohomology in the context of local supersymmetry. In this respect, an interesting problem would be to study the $AdS_2$ path integral from the world-sheet point of view using localization techniques because in this case there is a way to avoid the complications of local supersymmetry. This would be different from \cite{Beasley:2006us}, in the sense that we would be using  microcanonical boundary conditions instead.

Another interesting problem would be to extend this study to include also $\text{AdS}_2$ $\mathbb{Z}_c$ orbifolds. In \cite{Dabholkar:2014ema} it was shown how Chern-Simons theory on these orbifolds can reproduce the subleading Bessel functions in the $\CN=8$ microscopic answer. Namely, how topologically different orbifolds can reproduce the Kloosterman sums, which are intricate number theoretic objects.  However, a few key points of this computation are still missing. In particular, the convergence of the Rademacher expansion depends crucially on $c$, that is, the order of $\mathbb{Z}_c$. To see this we can construct a bound for each term in the Rademacher expansion \cite{Birmingham:2000xd}. The bound is of the order $ c^{\omega-3/2}$ , with $\omega$ the weight of the Jacobi form,  and thus the Rademacher expansion is convergent only for $\omega$ nonpositive (assuming that $\omega$ is half integer). It would be important to derive the exact dependence on the parameter $c$ at the cost of finding a divergent answer. 

On the same note, we can explore the contribution of the $\text{AdS}_2$ $\mathbb{Z}_c$ orbifolds on the background of mutiple instantons following \cite{Dabholkar:2014ema}. Given the Chern-Simons point of view explored in this work, in principle it would be straightforward to compute the Kloosterman sums for the $\CN=4$ theory and test them against a Jacobi-Rademacher expansion \cite{Dabholkar:2012nd}. This would be an additional test to the renormalization of the Chern-Simons levels that we proposed in section \S\ref{sec CS instantons}. 

Finally, it would be interesting to address the problem of small black holes and the DH-states in Heterotic string theory. They are certainly the simplest example concerning the microscopics. After all, we are counting only perturbative states. 

The partition function that captures the spectrum of half-BPS states is the modular form  
\begin{equation}
 \frac{1}{\eta(\tau)^{24}}=\sum_{n=-1}^{\infty}d(n)q^{n},
\end{equation}
and $d(n)$ is the degeneracy we want to study. In this case, the Rademacher expansion of $d(n)$ simplifies considerably when compared with the $\CN=8$ or the $\CN=4$ answers, and for this reason it is apparently more advantageous to compare with the bulk. The leading asymptotics is captured by the Bessel function 
\begin{equation}\label{1/2 BPS}
 d(n)\simeq I_{13}(4\pi\sqrt{n})\sim e^{4\pi\sqrt{n}-27/2\ln \sqrt{n}},\;n\gg 1
\end{equation}

From the bulk point of view, we expect  the Chern-Simons formulation to be still valid. However, the near horizon geometry preserves eight supercharges and hence we cannot use the $(0,4)$ Chern-Simons theory\footnote{By this we mean the Chern-Simons theory which has the same supersymmetry of the dual $(0,4)$ $CFT_2$.} to determine the exact measure as we did for the $\CN=4$ problem. Nevertheless, we can develop on the same idea and make a prediction for the contribution of different multiplets. The twenty two $\CN=4$ vector multiplets give a contribution of $-11\ln \sqrt{n}$ to the logarithmic correction and thus by (\ref{1/2 BPS}) we find a prediction for the supergravity multiplet of
\begin{equation}
 Z_{1\text{-loop}}^{\text{grav}}=(\sqrt{n})^{-5/2}.
\end{equation}
Part of it comes from integrating out the $SU(2)_L$ gauge fields. For the small black hole the full $S^3$ symmetry is restored\footnote{In this case we can set $e^1=0$ in the geometry (\ref{local AdS3xS3}) and thus we have $A^1=0$, modulo gauge transformations, in (\ref{5d CS reduction}).  } and thus the contribution of the $SU(2)_L$ part comes from fluctuations around the trivial connection $A_L=0$ which is in contrast with the black holes studied here. By the argument explained before, this leads to a contribution of $(\sqrt{n})^{-3/2}$ instead of $(\sqrt{n})^{-1/2}$, since we have now zero modes in all the three $su(2)_L$ directions. On the other hand the contribution from the $SL(2)_L$ part will still be the same which gives a mismatch of $(\sqrt{n})^{-1/2}$. It is possible that the contibution from the supersymmetric side does not cancel as we have more supersymmetry. It would be interesting to check this.

\subsection*{Acknowledgments}

We would like to thank Atish Dabholkar, Sameer Murthy and Valentin Reys for discussions on related topics and Alejandra Castro for comments on the draft. This work is part of the Delta ITP consortium, a program of the Netherlands Organisation for Scientific Research (NWO) that is funded by the Dutch Ministry of Education, Culture and Science (OCW). The author would also like to acknowledge DAMTP at the University of Cambridge where part of this work was done.

\bibliographystyle{JHEP}
\bibliography{measure2}

\end{document}